\documentclass[aps,amsfonts,reprint,tightenlines,amssymb,superscriptaddress,twocolumn, prapplied]{revtex4-2}  
\usepackage{graphicx}
\usepackage{dcolumn}
\usepackage{bm}
\usepackage{mathtools}
\usepackage{stmaryrd}
\usepackage{subcaption}
\usepackage{qcircuit}

\usepackage{caption}
\usepackage{ulem} 

\usepackage{algorithm}
\usepackage[noend]{algpseudocode}
\algrenewcommand\algorithmicdo{}

\makeatletter
\renewcommand{\ALG@name}{Procedure}
\makeatother

\usepackage[
  breaklinks=true,
  linktocpage=true,
  colorlinks=true, 
  pdfborder={0 0 0},
  linkcolor=blue,
  citecolor=red,
  filecolor=yellow,
  urlcolor=blue,
  bookmarks,
  pdfauthor={},
]{hyperref}

\usepackage{orcidlink}
\usepackage{xcolor}
\usepackage{pagecolor}
\usepackage{tabularx}
\usepackage{colortbl} 
\usepackage{xcolor}

\usepackage{ifthen}
\newcounter{is_qcircuit_used}
\newcommand{\red}[1]{}

\definecolor{myorange}{RGB}{255, 0, 0}

\begin{document}

\preprint{APS/123-QED}

\title{Quantum State Readout via Overlap-Based Feature Extraction}

\author{Hirofumi~Nishi\orcidlink{0000-0001-5155-6605}}
\email{hnishi@quemix.com}
\affiliation{
Quemix Inc.,
Taiyo Life Nihombashi Building,
2-11-2,
Nihombashi Chuo-ku, 
Tokyo 103-0027,
Japan
}
\affiliation{
Department of Physics,
The University of Tokyo,
Tokyo 113-0033,
Japan
}

\author{Taichi~Kosugi\orcidlink{0000-0003-3379-3361}}
\affiliation{
Quemix Inc.,
Taiyo Life Nihombashi Building,
2-11-2,
Nihombashi Chuo-ku, 
Tokyo 103-0027,
Japan
}
\affiliation{
Department of Physics,
The University of Tokyo,
Tokyo 113-0033,
Japan
}

\author{Xinchi~Huang\orcidlink{0000-0002-7547-074X}}
\affiliation{
Quemix Inc.,
Taiyo Life Nihombashi Building,
2-11-2,
Nihombashi Chuo-ku, 
Tokyo 103-0027,
Japan
}
\affiliation{
Department of Physics,
The University of Tokyo,
Tokyo 113-0033,
Japan
}

\author{Satoshi~Hirose\orcidlink{0000-0002-6576-9721}}
\affiliation{
Research Center of Excellence,
Honda R\&D Co., Ltd. \\
Shimotakanezawa 4630, Haga-machi, Haga-gun, 
Tochigi 321-3393, Japan
}

\author{Tatsuya~Okayama}
\affiliation{
Research Center of Excellence,
Honda R\&D Co., Ltd. \\
Shimotakanezawa 4630, Haga-machi, Haga-gun, 
Tochigi 321-3393, Japan
}

\author{Yu-ichiro~Matsushita\orcidlink{0000-0002-9254-5918}}
\affiliation{
Department of Physics,
The University of Tokyo,
Tokyo 113-0033,
Japan
}
\affiliation{
Quemix Inc.,
Taiyo Life Nihombashi Building,
2-11-2,
Nihombashi Chuo-ku, 
Tokyo 103-0027,
Japan
}
\affiliation{Quantum Materials and Applications Research Center,
National Institutes for Quantum Science and Technology (QST),
2-12-1 Ookayama, Meguro-ku, Tokyo 152-8550, Japan
}

\date{\today}

\begin{abstract}
In this study, a method for quantum state readout and feature extraction is developed using quantum-overlap-based fitting of function expansions. The approach involves the quantum calculation of quantum overlaps between a target quantum state and a linear combination of non-orthogonal basis functions, such as Lorentzian functions, via measurements, and classical optimization of the parameters in the function expansion. This method is particularly effective in scenarios where the quantum state is approximately represented as a combination of predefined basis functions. The proposed method involves a quantum state readout for both the complex amplitudes and their squared magnitudes (probabilities) in the quantum state. Numerical simulations were performed to reconstruct the grid-based wave function and X-ray absorption spectra from a quantum state, and the results show that our proposed method requires fewer measurements compared to conventional quantum state measurement techniques.
\end{abstract}

\maketitle

\section{INTRODUCTION}

Quantum state tomography (QST) has been the subject of extensive research for several years \cite{Vogel1989PRA, Hradil1997PRA, Pierre1999JPA, Dodonov1997PLA, James2001PRA, Blume2010NJP, Gross2010PRL, Cramer2010Ncom, Zhen2024IEEE, Guo2024CommPhys, Teng2025PRXQ}. It is regarded as a fundamental method for reconstructing the full quantum state of quantum systems. Conventional approaches require a substantial number of measurements and data analyses for each qubit in the system, giving rise to challenges concerning the computational burden and efficiency of measurement resources. In light of these challenges, researchers have proposed advanced statistical methods including maximum likelihood estimation \cite{James2001PRA}, Bayesian estimation \cite{Blume2010NJP}, and compressed sensing \cite{Gross2010PRL}. These methods have been incorporated into quantum state tomography, leading to significant improvements in accuracy and the ability to overcome the limitations of conventional methods.

However, in many quantum-computing applications, it is not necessary to reconstruct the full quantum state. Instead, it is often sufficient to estimate specific observables or physical quantities.
For example, energy estimation of a second-quantized Hamiltonian requires only $O(n_o^4)$ measurements, where $n_o$ is the number of spin-orbitals. Additional savings arise from grouping Pauli operators into commuting sets \cite{Bonet2020PRX, Hamamura2020npjQI, Huggins2022PRL, Inoue2024PRRes}. 
Related approaches, including classical shadows \cite{Aaronson2018ACM, Huang2020NPhys, Huang2021PRL, Hadfield2022CMP, Akhtar2023Quantum, Grier2024Quantum, West2025JPAMT} and quantum machine learning \cite{Cong2019NPhys, Pesah2021PRX, Budiutama2024PRA, Williams2024arXiv}, have been developed to extract relevant properties or features of quantum systems without explicitly reconstructing the full quantum state. These developments highlight that full-state reconstruction is not always the most efficient or necessary objective.

However, there are important applications in which the full reconstruction of the quantum state is essential. In computational physics applications relevant to computer-aided engineering (CAE) and computational fluid dynamics (CFD), such as electron dynamics, flow simulations, and Poisson equation solvers, the quantum state often directly encodes the solution on a spatial grid, and its full reconstruction is required to visualize or further process the solution \cite{Harrow2009PRL, Cao2013NJP, Berry2014JPA, Liu2021PNAS, Huang2024arXiv, Kassal2008PNAS, Kosugi2022PRR}. It is also required in self-consistent-field (SCF) calculations for nonlinear differential equations, where the reconstructed quantum state serves as input for the next iteration \cite{Nishiya2024arXiv}.
Therefore, for such CAE/CFD-related applications, the central task is not merely to estimate selected observables but to reduce the measurement cost of the full quantum state readout.

From a different perspective, a method for efficiently encoding quantum states by making assumptions regarding the shape of the quantum state was proposed \cite{Kosugi2024PRA, Kosugi2025PRA}. Refs.~\onlinecite{Kosugi2024PRA, Kosugi2025PRA} proposed an encoding method for spatially localized molecular orbitals, under the situation that the first-quantized Hamiltonian is simulated in quantum chemistry calculations. Specifically, they developed a quantum circuit that achieves a linear combination of discrete Lorentzian function (LF) states by using a technique for a linear combination of unitary operators (LCU) \cite{Berry2015PRL, Kosugi2020PRRes} with $O(n\log n)$ depth regarding the number of qubits. This finding underscores the efficacy of formulating assumptions about quantum states as a means of developing efficient encoding methodologies.

In this study, we developed a method for quantum state readout using quantum-overlap-based fitting of function expansions. 
The primary goal of this study was to develop an efficient readout of real-valued pure quantum states that arise in CAE/CFD-related applications, where the target states are expected to have additional problem-dependent structures.
This is also relevant for the calculation of spectral functions using quantum phase estimation (QPE). 
This approach involves calculating the quantum overlaps between the target quantum state and a linear combination of basis functions, such as LF, through measurements. This method is particularly effective when the quantum state can be represented as a linear combination of predefined basis functions. Furthermore, it has been demonstrated that the implementation of quantum circuits becomes more efficient when only the square of the absolute value of the amplitude of the quantum state is needed.

One of the key distinctions of the present work from previous studies is the extension of the basis functions used for quantum-state readout to a non-orthogonal basis. By evaluating quantum overlaps, the authors of Refs.~\cite{Miyamoto2023QIP, Chelsea2023arXiv, Wu2025arXiv} proposed readout methods to obtain the coefficients of orthogonal/orthonormal basis functions, such as Chebyshev polynomials or trigonometric functions. It should be noted that the evaluation of the weights of the basis functions at each grid point or for all basis functions based on quantum overlaps generally incurs an exponential computational cost. Miyamoto and Ueda alleviated the cost by approximating the coefficients as matrix-product states (MPS) \cite{Miyamoto2023QIP}. Besides, quantum Fourier series readout (FSR) schemes have been proposed to address these issues, in which the Fourier coefficients are directly observed by measuring the computational basis in Fourier space \cite{Huang2026QST, Huang2025arXiv2}. 
The FSR scheme automatically selects the dominant Fourier coefficient. 
In addition to previous studies using orthogonal basis functions, readout methods based on more efficient orthogonal polynomials have been explored \cite{Fang2026AMS}. 
In this approach, the state is reconstructed by maximizing the fidelity, where the optimization of the fidelity is employed using the LCU technique. 
In contrast, the proposed method emphasizes that the fidelity can be maximized without explicitly constructing a linear-combination state on the quantum circuit, which makes a major conceptual difference.

The remainder of this paper is organized as follows. Section~\ref{sect:method} introduces novel methodologies for reading quantum states using non-orthogonal basis function states considering two scenarios. The first scenario involves a method for calculating the quantum state, including the phase information, based on the fidelity score in Sect.~\ref{sect:fit_state}. The second scenario focuses on a method for calculating the squared absolute values of the coefficients on the computational basis of the quantum state based on the residual in Sect.~\ref{sect:fit}. In Sect.~\ref{sect:applications}, we apply the proposed method to a linear combination of localized function states and the probability distribution generated by QPE. Section~\ref{sect:numerical_results} presents numerical evidence demonstrating the efficacy of the proposed scheme in successfully reading out quantum states. 
In Sect.~\ref{sec:discussions}, we discuss the applicability of the proposed method to larger problems and generic highly entangled states, and compare it with existing tomography methods and variational quantum algorithms.
Finally, Section \ref{sect:conclusions} summarizes the findings and conclusions.

\section{METHOD}
\label{sect:method}

\begin{figure*}[ht]
    \centering
    \includegraphics[width=0.95 \textwidth]{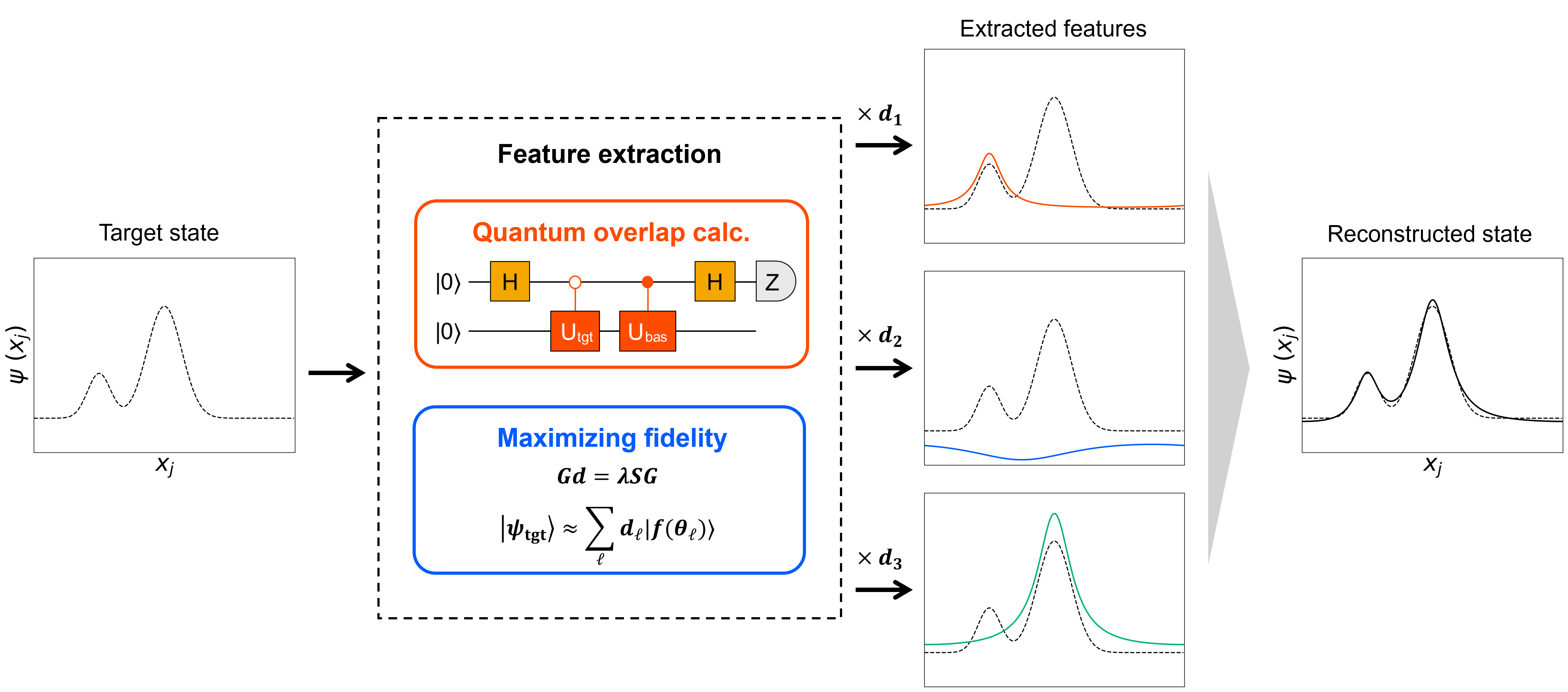} 
\caption{
Schematic of the proposed quantum-classical hybrid readout methodology. The quantum overlaps between a set of basis function states and the target quantum state are computed using a quantum computer. The resulting overlap data are used in a classical optimization procedure to maximize fidelity, thereby extracting the features of the target state. The target quantum state is expressed as a linear combination of basis function states.
}
\label{fig:method}
\end{figure*}

The overall procedure of the proposed quantum-classical hybrid readout method is illustrated in Fig.~\ref{fig:method}. In this approach, the inner products (alternatively called quantum overlaps) between the target quantum state and a set of predefined basis function states are evaluated using a quantum processor. These overlap values are then fed into a classical optimization routine that maximizes the fidelity between the reconstructed and target states. Through this optimization, the key features of the target state are extracted, and the state is ultimately represented as a linear combination of the basis function states.

\subsection{Readout of quantum states}
\label{sect:fit_state}
We consider the readout of an unknown target quantum state for an $n=\log_{2}N$ qubits system 
\begin{gather}
    |\psi_{\mathrm{tgt}}\rangle
    =
    \sum_{k=0}^{N-1} c_k |k\rangle_{n}
\label{eq:target_state}
\end{gather} 
generated by a unitary gate $U_{\mathrm{tgt}}$. Here, $|k\rangle_{n}$ denotes computational basis of the $n$-qubit system, while $c_k$ is its expansion coefficient. In this paper, we assume that 
the target quantum state is well approximated by a linear combination of $n_{\mathrm{bas}}$ non-orthogonal basis function (LCNOF) states $|f(\boldsymbol{\theta}_{\ell})\rangle$, given by
\begin{gather}
    |\psi_{\mathrm{LCNOF}} (\boldsymbol{d}, \boldsymbol{\theta}) \rangle
    =
    \sum_{\ell=0}^{n_{\mathrm{bas}}-1}
    d_{\ell} |f(\boldsymbol{\theta}_{\ell})\rangle.
\label{eq:mlf_state}
\end{gather}
The basis function state $|f (\boldsymbol{\theta}_{\ell})\rangle$ could be parameterized by $\boldsymbol{\theta}$ and generated by a unitary gate $U_{f (\bm{\theta}_{\ell})}$. 
The non-orthogonal basis function state includes discrete Lorentzian function state \cite{Klco2020PRA} and normal distribution state \cite{Iaconis2024npjQI, Manabe2025Quantum}, where $\theta$ can be the parameters to indicate the peak location and the variance of the states.
The readout of the target quantum state is achieved by maximizing the overlap between $|\psi_{\mathrm{tgt}}\rangle$ and $|\psi_{\mathrm{LCNOF}}\rangle$ with respect to the parameters, $\boldsymbol{d}, \boldsymbol{\theta}$:
\begin{gather}
    F (\boldsymbol{d}, \boldsymbol{\theta})
    \equiv 
    |
        \langle \psi_{\mathrm{tgt}}|
        \psi_{\mathrm{LCNOF}}(\boldsymbol{d}, 
        \boldsymbol{\theta})\rangle
    |^2,
\end{gather}
subject to 
$
    \langle \psi_{\mathrm{LCNOF}}(\boldsymbol{d}, 
    \boldsymbol{\theta})|
    \psi_{\mathrm{LCNOF}}(\boldsymbol{d}, 
    \boldsymbol{\theta})\rangle
     = 1
$.
This implies that parameters $\boldsymbol{d}$ and $\boldsymbol{\theta}$ function as features of the target quantum state.

First, we determine the optimal coefficients $\widetilde{\boldsymbol{d}}$ for a fixed $\boldsymbol{\theta}$, which means $\bm{\theta}_\ell$ is fixed for each $\ell=0,\ldots,n_{bas}-1$. The stationary condition for the overlap with respect to the parameter $\boldsymbol{d}$ is given by 
\begin{align}
    &
    \left(
        \frac{\partial F (\boldsymbol{d}, \boldsymbol{\theta})}{\partial d_{{\ell} }^{*}}
        -
        \lambda
        \frac{
            \partial
            \|
            | \psi_{\mathrm{LCNOF}} (\boldsymbol{d}, \boldsymbol{\theta}) \rangle
            \|^2
            }{\partial d_{{\ell}}^{*} }
    \right)_{\boldsymbol{d} = \widetilde{\boldsymbol{d}} }
    \nonumber \\
    &=
        \sum_{{\ell}^{\prime}}
        \left(
            G_{{\ell}, {\ell}'}
            ( \boldsymbol{\theta})     
            -
            \lambda
            S_{{\ell}, {\ell}^{\prime}} 
            ( \boldsymbol{\theta})
        \right)
            \widetilde{d}_{{\ell}^{\prime} }=0
        ,
    \label{ampl_encoding_of_GMO:stationarity_cond_for_d}
\end{align}
where the Hermitian matrix $G ( \boldsymbol{\theta})$ is defined as 
\begin{align}
    G_{{\ell}, {\ell}'}
    ( \boldsymbol{\theta})
    \equiv
        b_{\ell}^{*}
        ( \boldsymbol{\theta} )
        b_{\ell'}
        ( \boldsymbol{\theta})
        ,
    \label{ampl_encoding_of_GMO:def_G_mat}
\end{align}
with 
\begin{gather}
    b_{\ell} (\boldsymbol{\theta})
    \equiv
    \langle \psi_{\mathrm{tgt}} | f(\boldsymbol{\theta}_{\ell}) \rangle.
\end{gather}
$\lambda$ denotes the Lagrange multiplier for respecting the normalization condition and $S(\boldsymbol{\theta})$ is the overlap matrix between two non-orthogonal basis function states given by
\begin{gather}
    S_{\ell, \ell^{\prime}}
    (\boldsymbol{\theta})
    =
    \langle f (\boldsymbol{\theta}_{\ell}) | f (\boldsymbol{\theta}_{\ell'}) \rangle.
\label{eq:state_overlap}
\end{gather}
The stationary condition in
Eq.~(\ref{ampl_encoding_of_GMO:stationarity_cond_for_d})
can be rewritten as the generalized eigenvalue problem, given by
\begin{align}
    G ( \boldsymbol{\theta} )
    \widetilde{\boldsymbol{d}}
    =
        \lambda
        S ( \boldsymbol{\theta} )
        \widetilde{\boldsymbol{d}}
        .
\label{ampl_encoding_of_GMO:eig_prob_for_lc_coeffs}
\end{align}
We find that the eigenvector $\widetilde{\boldsymbol{d}}$ corresponding to the largest eigenvalue $\lambda_{\max} \equiv \max \lambda$ gives the highest fidelity, i.e.
$
    \lambda_{\max} (\bm{\theta})
    =
    F
        (
            \widetilde{\boldsymbol{d}},
            \boldsymbol{\theta}
        )
$ \cite{Kosugi2025PRA}.

Here, we consider the construction of $G(\boldsymbol{\theta})$ and $S(\boldsymbol{\theta})$ defined above. The overlap matrix for known functions $S(\boldsymbol{\theta})$ may be efficiently evaluated on a classical computer. 
For discrete LF states, the analytical expression of the overlap matrix can be obtained from Eq.~(\ref{eq:overlap_two_lfs}) in Appendix~\ref{sect:appendix_overlap}. 
As another example, when quantum states are represented in the MPS form with a small bond dimension \cite{Iaconis2024npjQI, Manabe2025Quantum}, the inner product between two such states can be evaluated efficiently.
However, the inner product $b_{\ell}(\boldsymbol{\theta})$ between the target state and the non-orthogonal function state must be evaluated on a quantum computer because the target quantum state is unknown. This evaluation is performed using the SWITCH test \cite{Buhrman2001PRL, Chamorro2023JPAMT}, as depicted in Fig.~\ref{circuit:switch_test}. 
The probability of observing the ancilla qubit as $|j\rangle$ is given by
\begin{gather}
    P_j
    =
    \frac{
        1 + (-1)^j \operatorname{Re} 
        e^{i\phi} 
        \langle \psi_{\mathrm{tgt}}
        |f(\bm{\theta}_{\ell})\rangle
    }{2} .
\label{eq:prob_switch_test}
\end{gather}
By changing the angle parameter $\phi$ of the phase gate in the SWITCH test, the real and imaginary parts of the inner product can be obtained.


\begin{figure}[h]
\centering
\mbox{ 
\Qcircuit @C=1em @R=1.6em{
\lstick{|0\rangle}             & \qw                     & \gate{H} & \gate{Z({\phi})} & \ctrlo{1}   & \ctrl{1}   &  \gate{H}  &  \meter \\
\lstick{|0\rangle_{n}} & \ustick{\otimes {n}} \qw  & {/} \qw  & \qw & \gate{U_\mathrm{tgt}} & \gate{U_{f(\bm{\theta}_{\ell})}} &  \qw       & \qw \\
}
} 
\caption{
Quantum circuit for calculating inner product 
$
    \operatorname{Re}e^{i\phi} 
    \langle \psi_{\mathrm{tgt}}
        |f(\bm{\theta}_{\ell})\rangle
$ in Eq.~(\ref{eq:prob_switch_test}) using SWITCH test.}
\label{circuit:switch_test}
\end{figure}

Next, we can maximize the reduced fidelity
$
    F(\boldsymbol{\theta})
    \equiv
    F(\widetilde{\boldsymbol{d}}, \boldsymbol{\theta})
$ with respect to the basis-function parameters $\boldsymbol{\theta}$ by any gradient-based optimization scheme.
The gradient with respect to $\boldsymbol{\theta}$ is given by \cite{Kosugi2025PRA} 
\begin{gather}
    \frac{
        \partial F(\boldsymbol{\theta})
    }{
        \partial {\bm{\theta}_{\ell}}
    }
    =
    \widetilde{\bm{d}}^{\dagger}
    \left(
        \frac{
            \partial G
            (\boldsymbol{\theta})
        }{\partial \bm{\theta}_{\ell}}
        -
        \lambda_{\max}
        \frac{
            \partial S(\boldsymbol{\theta})
        }{\partial \bm{\theta}_{\ell}}
    \right)
    \widetilde{\bm{d}}.
\end{gather}
Analogous to the construction of the generalized eigenvalue problem, the derivative of the overlap matrix $\partial_{\theta_{\ell}} S(\boldsymbol{\theta})$ is calculated using a classical computer. The derivative of the quantum overlap $\partial_{\theta_{\ell}} G(\boldsymbol{\theta})$ must be evaluated on a quantum computer; however, the circuit implementation for this derivative is not clear. Consequently, we opted to implement a finite-differentiation method to evaluate the derivative using a quantum computer.

In case of discrete variables of $\boldsymbol{\theta}$, on the other hand, we can employ the Metropolis method for the optimization.
Specifically, we determine whether or not to update the position with a probability according to the Boltzmann distribution, given by
$
\min\{1, \exp(\beta\Delta F)\}
$, 
where 
$
    \Delta F 
    = 
    F(\boldsymbol{\theta}^{(n)})
    -
    F(\boldsymbol{\theta}^{(n-1)})
$ with $n$-step centers  $\boldsymbol{\theta}^{(n)}$. 
$\beta$ is a parameter called the inverse temperature. 

We stress that the readout method proposed here does not require explicit preparation of a linear combination of localized functions in contrast to the amplitude-encoding method \cite{Kosugi2024PRA, Kosugi2025PRA}. In amplitude encoding, the success probability depends on the number of bases of the localized function $O(1/n_{\mathrm{bas}})$ because it is necessary to implement a linear combination of unitary operators. However, in the readout case, the problem related to success probability does not appear because  the linear combination of the localized function is considered as classical post-processing. However, it is necessary to compute the quantum overlaps involving the unknown quantum state, which are required for parameter optimization using quantum computation. A detailed discussion of the computational cost associated with this calculation can be found in Sect.~\ref{sect:fit_cost}.

\subsection{Readout of  amplitude of quantum states}
\label{sect:fit}
In this subsection, we consider the estimation of the absolute square of the coefficients of the computational basis $\boldsymbol{y}_{\mathrm{tgt}} = (|c_0|^2, |c_1|^2, \ldots, |c_{N-1}|^2)$, in contrast to the scenario for the readout of the quantum states, including the phase factor in Sect.~\ref{sect:fit_state}.
Let us consider an $N$-component vector as a linear combination of the discrete non-orthogonal basis functions expressed as 
\begin{gather}
    \boldsymbol{y}_{\mathrm{LCNOF}}(
        \boldsymbol{d},
        \boldsymbol{\theta}
    )
    =
    \sum_{\ell=0}^{n_{\mathrm{bas}}-1} 
    d_{\ell} \boldsymbol{y}(\boldsymbol{\theta}_{\ell}),
\label{eq:mlf_y}
\end{gather}
where $
    y_{k}(\boldsymbol{\theta}_{\ell})
    =
    |f_{k}(\boldsymbol{\theta}_{\ell})|^2 
    = 
    |\langle k |f(\boldsymbol{\theta}_{\ell})\rangle|^2
$.
First, we optimize the coefficients $\boldsymbol{d}$ for a fixed $\boldsymbol{\theta}$. The optimal coefficients $\widetilde{\boldsymbol{d}}(\boldsymbol{\theta})$ are derived by minimizing the squared norm:
\begin{gather}
    \widetilde{\boldsymbol{d}}
    =
    \arg\min_{\boldsymbol{d}} 
    \left\| 
        \boldsymbol{y}_{\mathrm{tgt}} 
        -
        \sum_{\ell=0}^{n_{\mathrm{bas}}-1}
        d_{\ell} \boldsymbol{y}(\boldsymbol{\theta}_{\ell})
    \right \|^2.
\label{eq:min_for_optimal_d}
\end{gather}
The optimization problem Eq.~(\ref{eq:min_for_optimal_d}) is equivalent to the following equation
\begin{gather}
    Q (\boldsymbol{\theta}) \widetilde{\boldsymbol{d}}
    =
    \boldsymbol{h} (\boldsymbol{\theta})
\label{eq:lineq_for_optimal_d}
\end{gather}
where 
\begin{gather}
    Q_{\ell, \ell^{\prime}}(
        \boldsymbol{\theta}
    )
    \equiv
    \sum_{k=0}^{N-1}
    |f_{k}(\boldsymbol{\theta_{\ell}})|^2
    |f_{k}(\boldsymbol{\theta_{\ell^{'}}})|^2
\label{eq:overlap_matrix}
\end{gather}
and 
\begin{gather}
\begin{aligned}
    h_{\ell}(\boldsymbol{\theta})
    \equiv 
    \sum_{k=0}^{N-1}
    |c_k|^2 |f_{k}(\boldsymbol{\theta}_{\ell})|^2
    .
\label{eq:h_vector}
\end{aligned}
\end{gather}
The overlap matrix between the squared basis function in Eq.~(\ref{eq:overlap_matrix}) is calculated using a classical computer, as mentioned in Sect.~\ref{sect:fit_state}.
The quantity $h_{\ell}(\boldsymbol{\theta})$ in Eq.~(\ref{eq:h_vector})
can be evaluated using a SWAP test \cite{Buhrman2001PRL} in Fig.~\ref{circuit:swap_test}.
We first prepare the $2n$-qubit state
\begin{gather}
    |\psi_{\mathrm{tgt}}^{\prime}\rangle
    =
    \sum_{k=0}^{N-1}
    c_k
    |k\rangle_{n}\otimes|k\rangle_{n},
\label{eq:target_swap}
\end{gather}
from $|\psi_{\mathrm{tgt}}\rangle$ using CNOT gates.
Tracing out either one of the two registers gives
\begin{gather}
    \rho_{\mathrm{tgt}}^{\mathrm{diag}}
    =
    \sum_{k=0}^{N-1}
    |c_k|^2 |k\rangle\langle k|.
\label{eq:target_diag_density}
\end{gather}
Therefore,
\begin{gather}
    h_{\ell}(\boldsymbol{\theta})
    =
    \langle f(\boldsymbol{\theta}_{\ell})|
    \rho_{\mathrm{tgt}}^{\mathrm{diag}}
    |f(\boldsymbol{\theta}_{\ell})\rangle .
\label{eq:value_by_swap_test}
\end{gather}
The probability of observing the ancilla qubit in the state $|j\rangle$
is then given by
\begin{gather}
    P_j
    =
    \frac{
        1+(-1)^j h_{\ell}(\boldsymbol{\theta})
    }{2}.
\label{eq:prob_swap_test}
\end{gather}
Here, the number of qubits in the target quantum state and the basis function states does not necessarily have to match. If they differ, the probability distribution of the larger-qubit state is marginalized over its least significant qubits. See Appendix~\ref{sect:overlap_diff_qubits} for details.

\begin{figure}[h]
\centering
\mbox{ 
\Qcircuit @C=1em @R=1.6em{
\lstick{|0\rangle}     & \qw                     & \qw      & \gate{H}               & \ctrl{2} & \gate{H} & \meter \\
\lstick{|0\rangle_{n}} & \ustick{\otimes n} \qw  & {/} \qw  & \gate{U_{f(\bm{\theta}_{\ell})}} & \qswap   & \qw & \qw \\
\lstick{|0\rangle_{n}} & \ustick{\otimes n} \qw  & {/} \qw  & \targ                    & \qswap   & \qw & \qw \\
\lstick{|\psi_{\mathrm{tgt}}\rangle} & \ustick{\otimes n} \qw  & {/} \qw  & \ctrl{-1}  & \qw      & \qw & \qw \\
}
} 
\caption{
The quantum circuit used to evaluate overlap $h_{\ell}(\boldsymbol{\theta})$ in Eq.~(\ref{eq:value_by_swap_test}) using the SWAP test. 
}
\label{circuit:swap_test}
\end{figure}

Solving Eq.~(\ref{eq:lineq_for_optimal_d}) leads to the optimal coefficients as
\begin{gather}
    \widetilde{\boldsymbol{d}}
    =
    Q^{-1}(\boldsymbol{\theta})
    \boldsymbol{h}(\boldsymbol{\theta})
    .
\end{gather}
The residual norm is evaluated with
\begin{gather}
    \|
    \boldsymbol{r} (\boldsymbol{\theta})
    \|^2
    =
    \boldsymbol{y}_{\mathrm{tgt}}\cdot 
    \boldsymbol{y}_{\mathrm{tgt}} 
    -
    \widetilde{\boldsymbol{d}}
    \cdot
    Q(\boldsymbol{\theta})
    \widetilde{\boldsymbol{d}}
    .
\end{gather}
The first term on the RHS is evaluated using the SWAP test, where both inputs are the target states. 
For the optimization problem itself, it is not necessary to calculate the first term as it is a constant.
The optimization of the basis-function parameters $\boldsymbol{\theta}$ is the same as those in Sect.~\ref{sect:fit_state}. The derivative of the residual norm with respect to $\theta_{\ell}$ is derived as
\begin{gather}
    \frac{\partial}{\partial \bm{\theta}_{\ell}}
    \|
    \boldsymbol{r} (\boldsymbol{\theta})
    \|^2
    =
    -2\widetilde{\bm{d}} \cdot \frac{\partial \bm{h}(\bm{\theta})}{\partial \bm{\theta}_\ell} 
    + 
    \widetilde{\bm{d}} \cdot \frac{\partial Q(\bm{\theta})}{\partial \bm{\theta}_\ell} 
    \widetilde{\bm d}
    .
\end{gather}
The derivative $\partial_{\theta_{\ell}} \boldsymbol{h} (\bm{\theta})$ is evaluated using a finite difference.

The following discussion addresses the differences between the methods proposed in Sect.~\ref{sect:fit_state} and the proposed method in this section. The method outlined in Sect.~\ref{sect:fit_state} involves the extraction of information including the phase of the quantum state. By contrast, the method proposed in this section is confined to the calculation of the absolute square of the coefficients. This limitation allows the quantum circuit to simplify the calculation of the inner product in a SWAP test rather than a SWITCH test. The SWITCH test necessitates the incorporation of an overhead in the quantum circuit, owing to the requirement of a control gate to generate the target state. 
We note, however, that the SWAP test can also be used within the framework of Sect.~\ref{sect:fit_state} when the target amplitudes are real and non-negative up to an overall phase factor and the basis functions are non-negative, since $b_{\ell}(\boldsymbol{\theta})$ is then recovered as 
$
    \sqrt{|\langle \psi_{\mathrm{tgt}}|f(\boldsymbol{\theta}_{\ell})\rangle|^2}
$.
In such a case it is advantageous to use the method described in Sect.~\ref{sect:fit_state}, because the phase information is unnecessary and the quantum overlap can therefore be evaluated by a SWAP test, while, unlike the method presented in this section, the normalization of the reconstructed state is preserved.

\subsection{Amplitude estimation}
\label{sect:fit_qae}
As the quantum state is obtained probabilistically through observation, numerous repetitions are necessary, and the scaling of the error follows Monte Carlo sampling: $\varepsilon = O(1/\sqrt{N_{\mathrm{shot}}})$ where $N_{\mathrm{shot}}$ is the number of measurements.
The scaling of the error can be improved quadratically by leveraging quantum amplitude estimation (QAE) techniques \cite{Brassard2002}. 
This section explores the application of the QAE to estimate the quantum overlaps, as mentioned in previous sections.

Amplitude amplification (AA) operator is given by 
\begin{gather}
    Q
    =
    - \mathcal{U} S_{0} \mathcal{U}^{\dagger} S_{\chi}
    ,
\label{eq:aa_operator}
\end{gather}
where $S_{\chi}$ and $S_0$ denote oracle and zero reflection, respectively. 
$\mathcal{U}$ represents the unitary operator to generate a quantum state to be estimated.
The zero reflection $S_0$ is implemented with $O(n)$ two-qubit gates with a single ancilla qubit \cite{AdrianoPRA1995, MaslovPRA2016}. 
In the SWITCH and SWAP tests, we want to estimate the observation probability of the single ancilla qubit, so the oracle is represented as $S_{\chi} = Z $, which is a single quantum gate acting on the ancilla qubit.
\begin{figure}[ht]
\centering
\mbox{ 
\Qcircuit @C=1em @R=1.4em{
& \gate{S_{\chi}}  & \gate{H} & \ctrl{2} &  \gate{H} & \qw & \multigate{3}{S_0} 
& \qw                   & \gate{H}  & \ctrl{2} & \gate{H} & \qw
\\
& \ustick{\otimes n} \qw  & {/} \qw & \qswap & \qw & \gate{U_{f(\bm{\theta_{\ell}})}^{\dagger}} &\ghost{S_0}
& \gate{U_{f(\bm{\theta_{\ell}})}}  & \qw       & \qswap   & \qw   & \qw    
\\
& \ustick{\otimes n} \qw  & {/} \qw & \qswap & \targ & \qw &\ghost{S_0}
& \qw                   & \targ     & \qswap   & \qw       & \qw
\\
& \ustick{\otimes n} \qw  & {/} \qw & \qw    & \ctrl{-1} & \gate{U_{\text{tgt}}^{\dagger}} &\ghost{S_0}
& \gate{U_{\text{tgt}}} & \ctrl{-1} & \qw      & \qw       & \qw 
\\
}
} 
\caption{
Quantum circuit for the AA operator in Eq.~(\ref{eq:aa_operator}) using the SWAP test. 
}
\label{fig:qc_qae_swap}
\end{figure}

Let $|\psi_{\text{good}}\rangle$ be the state we want to amplify, and $|\psi_{\text{bad}}\rangle$ be the state that is orthogonal to it, satisfying $\langle \psi_{\text{good}} |\psi_{\text{good}}\rangle = a$ and $\langle \psi_{\text{bad}} |\psi_{\text{bad}}\rangle = 1-a$.
In this study, the ancilla qubits of the SWITCH test or SWAP test are considered to be in the good state $|\psi_{\text{good}}\rangle$ when the ancilla state in the $|0\rangle$ state and in the bad state $|\psi_{\text{bad}}\rangle$ when the ancilla state in the $|1\rangle$ state.
The AA operator satisfies
\begin{gather}
    Q|\psi_{\pm}\rangle
    =
    \lambda_{\pm} |\psi_{\pm}\rangle,
\end{gather}
where 
\begin{gather}
   |\psi_{\pm}\rangle
   =
   \frac{1}{\sqrt{2a}}
   |\psi_{\mathrm{good}}\rangle
   \pm
   \frac{i}{\sqrt{2(1-a)}} 
   |\psi_{\mathrm{bad}}\rangle
\end{gather}
and $\lambda_{\pm} = e^{\pm 2i \theta_a}$ with $\sin^2 \theta_a = a$.
Assuming that the AA is the unitary operator of QPE, the amplitude of the AA operator can be estimated using QPE via the estimation of the phase of the AA operator.
The quantum circuit for the AA operator that is tasked with the SWAP test is depicted in Fig.~\ref{fig:qc_qae_swap}. It is imperative to note that the original QAE necessitates the incorporation of a control operation for the AA operator. Hence, the addition of the control operations for the oracle and zero reflection is necessary. A proposal to circumvent the necessity for additional control operations is put forth in \cite{Suzuki2020QIP, Tanaka2021QIP, Grinko2021npjQI, Giurgica2022Quantum, Manzano2023EPJ, Rall2023Quantum, Ghosh2024IEEE}, wherein the amplitude is estimated through results obtained by multiple repetitions of the AA operators.

In the original QAE, the circuit depth increases as
$
    O\left(
        1 / \varepsilon 
    \right),
$
however the measurement of the ancilla qubits takes $O(1)$ repetitions.
Therefore, we achieve the Heisenberg-limited scaling with respect to precision $\varepsilon$.

\subsection{Computational cost}
\label{sect:fit_cost}
We consider the classical computational cost of this measurement protocol.  The construction of the overlap matrix requires $O(n_{\mathrm{bas}}^2 M)$, where $M$ denotes the computational cost of the overlap integral.
In this work, we assume that non-orthogonal basis function states can be prepared efficiently. In many practical settings, the inner products between such basis functions can also be computed efficiently. For example, when quantum states are prepared using MPS encodings \cite{Iaconis2024npjQI, Manabe2025Quantum}, the inner product between two MPS states can be evaluated rapidly. Moreover, Lorentzian functions correspond to rank-1 MPS representations.
Solving the generalized eigenvalue equation or linear equation requires $O(n_{\mathrm{bas}}^3)$. For optimizing the parameters within the basis function states, this classical computation is iterated until the fidelity or residual norm converges. Let us denote the number of iterations by $n_{\mathrm{iter}}$, the total classical computational cost is $c_{C} = O(n_{\mathrm{iter}}\max(n_{\mathrm{bas}}^2 M, n_{\mathrm{bas}}^3)) $.

We consider the computational cost required for quantum computing. Let the computational cost of preparing the target state be $c_{\text{tgt}}$. In this measurement process, the total computational cost is $O(c_{\text{tgt}} m_{\mathrm{eval}} N_{\mathrm{shot}})$, where the number of evaluations for the quantum overlaps is denoted by $m_{\mathrm{eval}}$ and  $m_{\mathrm{eval}} \leq \min(n_{\mathrm{iter}} n_{\mathrm{bas}}, N_\theta)$.
Here, $N_\theta$ denotes the number of elements of the admissible set for $\mathbf{\theta}$. 
(For example, $N_\theta=N$ if $\theta$ is a scalar and takes the grid size $N$ (i.e., number of grid points)).
 
We assumed that the computational cost required for the SWAP and SWITCH tests was smaller than $c_{\text{tgt}}$, and therefore ignored it. 
The required number of measurements $N_{\text{shot}}$ depends on the infidelity $\delta$, which is given as
$
    \delta 
    = 
    O(n_{\mathrm{bas}}\kappa(S)^2 /(N_{\mathrm{shot}} \Delta\lambda^2 ))
$ for the generalized eigenvalue equation in Eq.~(\ref{ampl_encoding_of_GMO:eig_prob_for_lc_coeffs}).
$\kappa(S)$ denotes the condition number of the overlap matrix $S$.
$\Delta \lambda$ is the spectral gap, $\Delta\lambda = \lambda_0 - \lambda_1 $, in Eq.~(\ref{ampl_encoding_of_GMO:eig_prob_for_lc_coeffs}). 
See Appendix~\ref{sect:error_bound} for the details of the derivation. 
Finally, the quantum computational cost is estimated as 
\begin{gather}
        c_{Q}
    =
    O\left(
        \frac{
            c_{\mathrm{tgt}} m_{\mathrm{eval}}
            n_{\mathrm{bas}} \kappa(S)^2
        }{
            \delta \Delta\lambda^2
        }
    \right).
\end{gather}
These results suggest that the readout cost increases when the prepared basis functions do not adequately represent the target state or when the non-orthogonality of the overlap matrix is large.
We demonstrated that the utilization of the QAE could reduce the cost to $
    O(c_{\text{tgt}} m_{\mathrm{eval}}\sqrt{ n_{\mathrm{bas}} \kappa(S)^2/\delta\Delta\lambda^2})
$.

Compared with the conventional approach for estimating quantum states, the number of iterations $m_{\mathrm{eval}}$ serves as a critical metric for evaluating the efficacy of the proposed method. The number of iterations $m_{\mathrm{eval}}$ depends on  the initial parameters. The separation between initial and optimal parameters is directly correlated with the number of iterations $n_{\mathrm{iter}}$ and the number of quantum overlap evaluations $m_{\mathrm{eval}}$ using a quantum computer.

Nevertheless, it is possible to prepare an adequate initial state under certain circumstances. For instance, solutions employing mean-field approximation \cite{Hohenberg1964, Kohn1965} frequently yield suitable initial states. Moreover, the solutions from previous steps, such as SCF calculations, can also serve as adequate initial states. The initial parameter dependency was confirmed through numerical calculations in Sect.~\ref{sect:numerical_results}.

\section{Applications}
\label{sect:applications}
\subsection{Discrete Lorentzian function state}
For an $n$-qubit system, the discrete Slater function (SF) in one-dimensional (1D) space with $N\equiv 2^n$ elements is expressed as \cite{Klco2020PRA} 
\begin{gather}
    S_j (n,a)
    \equiv
    \begin{cases}
        {C}_S(n, a) e^{-a j}  
        &
        0 \leq j < N/2
        \\
        {C}_S(n, a) e^{-a(N- j)}
        &
        N/2 \leq j < N  
    \end{cases}
    ,
\label{eq:slater_function_state}
\end{gather}
where $a$ is the decay rate and ${C}_{S} (n, a)$ is the normalization constant, given by
\begin{gather}
    C_{S} (n, a)
    \equiv
    \sqrt{\frac{1 - e^{-2a}}{(1+e^{-2a}) (1-e^{-Na})}} .
\label{eq:norm_slater_function_state}
\end{gather}
The normalized SF state is defined as
$
    |S; a\rangle
    \equiv
    \sum_{k=0}^{N-1} S_k(n, a) |k \rangle_{n} 
$.
The discrete LF is given by \cite{Kosugi2024PRA}
\begin{align}
    L_k (n, a)
    \equiv
        \frac{{C}_{S} (n, a) }{\sqrt{N}}
        \frac{(1 - e^{-2 a}) (1 - (-1)^k e^{-a N/2})}
        { 1 - 2 e^{-a} \cos (2 \pi k/N) + e^{-2 a} }
        .
\label{ampl_encoding_of_GMO:def_discrete_Lorentzian}
\end{align}
Note that the normalized discrete LF is localized at the origin in 1D space. 
The origin-centered LF state can be shifted to a normalized LF state centered at an integer coordinate $k_{\mathrm{c}}$ as follows:
\begin{align}
    | L ; a, k_{\mathrm{c}} \rangle
    \equiv
        \sum_{k = 0}^{N - 1}
        L_{k - k_{\mathrm{c}}} (n, a)
        | k \rangle_{n}
    .
    \label{eq:def_Lorentzian_func_state}
\end{align}
The shifted LF state can be derived by applying the quantum Fourier transform (QFT) on the discrete SF state, i.e., $|L; a, k_{c}=0 \rangle = U_{\mathrm{QFT}} |S, a\rangle = U_{\mathrm{QFT}}^{\dagger} |S, a\rangle$. 
Translation operator is defined as 
$
    T(k) 
    \equiv 
    U_{\mathrm{QFT}}
    U_{\mathrm{shift}}(k) 
    U_{\mathrm{QFT}}^{\dagger}
$ \cite{Kosugi2024PRA},
which is confirmed to perform modular addition for a computational basis as  
$T(k)|j\rangle_{n} = |(j+k) \operatorname{mod} N\rangle_{n}$.
$U_{\mathrm{shift}}(k)$ is implemented with single-qubit phase gates, as explained in Appendix~\ref{sect:impl_discrete_lorentzian}. 
Then, the shifted LF state can be obtained by $|L; a, k_c\rangle = T(k_c)|L; a, 0\rangle$.
Also, the analytical expression for the overlap matrix is given in Appendix~\ref{sect:appendix_overlap}.

\subsection{X-ray absorption spectra}
\subsubsection{Review of QPE-sampling}
\label{sect:xas_qpe_sampling}
On a quantum computer, spectral functions for the many-body Hamiltonian $\hat{\mathcal{H}}$ can be calculated by employing QPE based on QFT \cite{Kosugi2020PRA, Kosugi2020PRRes}, as depicted in Fig.~\ref{circuit:qpe_sampling}.
\begin{figure}[h]
\centering
\mbox{ 
\Qcircuit @C=1em @R=1.2em{
\lstick{|0\rangle}
& \qw & \multigate{2}{U_{\mathrm{in}}} & \ctrl{3} & \qw & \qw & \multigate{2}{\text{QFT}^{\dagger}} & \meter \\
\lstick{|0\rangle}
& \qw & \ghost{U_{\mathrm{in}}}        & \qw & \ctrl{2} & \qw & \ghost{\text{QFT}^{\dagger}}        & \meter \\
\lstick{|0\rangle}
& \qw & \ghost{U_{\mathrm{in}}}        & \qw & \qw & \ctrl{1} & \ghost{\text{QFT}^{\dagger}}        & \meter \\
\lstick{|\Psi_{\mathrm{in}} \rangle} 
& \ustick{\otimes n_s} \qw   & {/} \qw   &\gate{U} & \gate{U^2} & \gate{U^4} &  \qw                      & \qw \\
}
} 
\caption{
Quantum circuit for the XAS spectra based on QPE-sampling when $n=3$. $U=e^{2\pi i t_0 \hat{\mathcal{H}} / N}$ is used in this figure.
}
\label{circuit:qpe_sampling}
\end{figure}
We consider the input state for an $n_s$-qubit system given by
\begin{gather}
    |\Psi_{\mathrm{in}}\rangle
    =
    \sum_{j=0}^{N_s-1}c_j |\Psi_{j}\rangle ,
\label{eq:qpe_input}
\end{gather}
where $N_s \equiv 2^{n_s}$. 
$|\Psi_{\mathrm{in}}\rangle$ is expanded with the eigenvectors of $\hat{\mathcal{H}}$, denoted by $|\Psi_j\rangle$, and $c_j$ denotes the expansion coefficients.
By operating the QPE circuit, the input state before measurement of the ancilla qubits is expressed as 
\begin{gather}
    \sum_{j=0}^{N_s-1} c_j 
    \sum_{k=0}^{N -1} 
    \alpha\left(t_0 E_j - k\right)
    |\Psi_{j}\rangle \otimes |k \rangle_{n}
    .
\label{eq:qpe_ouotput}
\end{gather}
$t_0$ represents a scaling parameter to increase or decrease the eigenvalues for precise measurement.
The function $\alpha$ depends on the unitary operator $U_{\mathrm{in}}$ in QPE.
In QPE, the possible energy eigenvalue $E_{j}$ is loaded to the $n$ ancilla qubits in the binary representation.
The probability of observing ancilla qubits as $|k\rangle_{n}$ state is given by
\begin{gather}
    P_k
    =
    \sum_{j=0}^{N_s-1} |c_j|^2
    \left|
        \alpha(t_0 E_j -k)
    \right|^2 .
\end{gather}

In this study, we choose $U_{\mathrm{in}}$ as the amplitude encoding for the discrete SF state \cite{Klco2020PRA, Fomichev2024arXiv}.
In this case, $\alpha$ is given by
\begin{gather}
    \alpha^{\mathrm{LF}}(x)
    =
    \frac{\widetilde{C}_S(n, a)}{\sqrt{N}}
    \sum_{\tau=0}^{N - 1}
    e^{-a\tau} e^{2\pi i \tau x/N} ,
\label{eq:qpe_alpha_lf}
\end{gather}
where $a$ denotes the decay rate of the discrete SF and the normalization constant
\begin{gather}
    \widetilde{C}_S(n, a) 
    = 
    \sqrt{\frac{1 - e^{-2a}}{1 - e^{-2aN}}} .
\label{eq:norm_const_mod}
\end{gather}
Since the Fourier transformation of the discrete SF is the discrete LF, the probability is approximated as 
\begin{gather}
    P_k
    =
    C_{\mathrm{const}} \eta
    \sum_{j=0}^{N_s-1} 
    \frac{|c_j|^2}{(t_0 E_j - k)^2 + \eta^2}
     ,
\label{eq:prob_dstb_qpe_lf}
\end{gather}
where $\eta = aN/(2\pi)$ and $N \gg 1$. $C_{\mathrm{const}}$ denotes a $k$-independent normalization factor.
The quantum circuit for $U_{\mathrm{in}}$ and the derivation of Eq.~(\ref{eq:qpe_alpha_lf}) are summarized in Appendix~\ref{sect:comparison_qpe}.
By selecting $U_{\mathrm{in}}$ in this way, the probability distribution obtained by QPE sampling naturally follows a shape with a LF broadening.
We denote this method by QPE-LF.
Such a distribution corresponds to the formulation of single-particle spectral functions \cite{Kosugi2020PRA} and X-ray absorption spectroscopy (XAS) spectra \cite{Kosugi2020PRRes, Fomichev2024arXiv, Sakuma2024PRA}.


\subsubsection{Lorentzian fitting of XAS spectra}
Next, we describe a method for estimating the probability distribution obtained by QPE in Eq.~(\ref{eq:prob_dstb_qpe_lf}) using the LF state.
Eq.~(\ref{eq:qpe_ouotput}) is rewritten as
\begin{gather}
    |\psi_{\mathrm{QPE}}\rangle
    =
    \sum_{k=0}^{N - 1}
    \widetilde{c}_k |\widetilde{\phi}_k \rangle
    \otimes |k \rangle_{n}
\label{eq:qpe_output_re}
\end{gather}
where 
\begin{gather}
    \widetilde{c}_k  |\widetilde{\phi}_k\rangle
    \equiv
    \sum_{j=0}^{N_s-1}
    c_j 
    \alpha\left(t_0 E_j - k\right)
    |\Psi_j\rangle .
\end{gather}
The absolute squared of the coefficients $\{|\widetilde{c}_k|^2 \}$ composes the spectral function obtained by QPE.

\begin{figure}[h]
\centering
\mbox{ 
\Qcircuit @C=1em @R=1.6em{
\lstick{|0\rangle}     & \qw                       & \qw      & \qw                       & \gate{H} & \ctrl{2} & \gate{H} &  \meter \\
\lstick{|0\rangle_{n}} & \ustick{\otimes n} \qw  & {/} \qw  & \gate{U_{\text{LF}}}  & \qw & \qswap & \qw & \qw \\
\lstick{|0\rangle_{n}} & \ustick{\otimes n} \qw  & {/} \qw  & \multigate{1}{\text{QPE}} & \qw & \qswap &  \qw & \qw \\
\lstick{|\Psi_{\mathrm{in}} \rangle} & \ustick{\otimes n_s} \qw  & {/} \qw  & \ghost{\text{QPE}}  & \qw  & \qw & \qw  & \qw \\
}
} 
\caption{
Quantum circuit for evaluating the overlap $|\langle L;a_{\ell}, k_{c\ell} | \psi_{\mathrm{QPE}}\rangle|^2$ in Eq.~(\ref{eq:h_value_QPE}) based on the SWAP test. 
}
\label{circuit:swap_test_qpe}
\end{figure}

Here, we also adopt the LCNOF state as a discrete LF state,
$|f(\theta_\ell)\rangle = \sum_{k=0}^{N-1} L_{k-k_{c\ell}}(n, a_\ell) |k\rangle$, where the parameters $\bm{\theta}_\ell = (k_{c\ell}, a_\ell)$.
Using quantum circuit drawn in Fig.~\ref{circuit:swap_test_qpe} provides the inner products for the linear equation in Eq.~(\ref{eq:lineq_for_optimal_d}) to optimize the coefficients for the LCNOF vector  in Eq.~(\ref{eq:mlf_y}):
\begin{gather}
\begin{aligned}
    h_{\ell}(a_{\ell}, k_{c\ell})
    &=
    |\langle L; a_{\ell}, k_{c\ell}
    | \psi_{\mathrm{QPE}}\rangle |^2
    \\ &=
    \sum_{k=0}^{N-1}
    |\widetilde{c}_k|^2 L^2_{k-k_{c\ell}}(n, a_{\ell})
    .
\label{eq:h_value_QPE}
\end{aligned}
\end{gather}
Here, we assume that the off-diagonal coherences between different computational-basis states of the QPE register are negligible.
We use the equivalent representation
$|L;a,k_c\rangle
=
U_{\mathrm{QFT}}^{\dagger}
U_{\mathrm{shift}}(-k_c)|S,a\rangle$
given in Appendix~\ref{sect:impl_discrete_lorentzian}.
In the overlap evaluation, the resulting QFT cancels the inverse QFT in the QPE circuit.

One may attempt to estimate the spectral function obtained by QPE sampling using a method based on the fidelity of Sect.~\ref{sect:fit_state}. However, the quantum circuit for calculating the inner product at this point becomes more complex than the method based on the residual depicted in Fig.~\ref{circuit:swap_test_qpe}.
The details are discussed in Appendix~\ref{sect:appendix_fidelity_based_qpe}.

We remark on the limitations of this method for readout of XAS spectra. As examined numerically in Appendix~\ref{sect:appendix_overlap}, the overlap matrix of the squared Lorentzian functions exhibits a normalization factor that decreases exponentially with the number of qubits. Although the computational cost presented in Appendix~\ref{sect:error_bound} corresponds to solving the generalized eigenvalue problem and is therefore somewhat different, the exponential decay of the overlap matrix may potentially increase the cost of the readout.

\section{Numerical results}
\label{sect:numerical_results}
\subsection{Readout of a quantum state}
\begin{figure*}[ht]
    \centering
    \includegraphics[width=0.9 \textwidth]{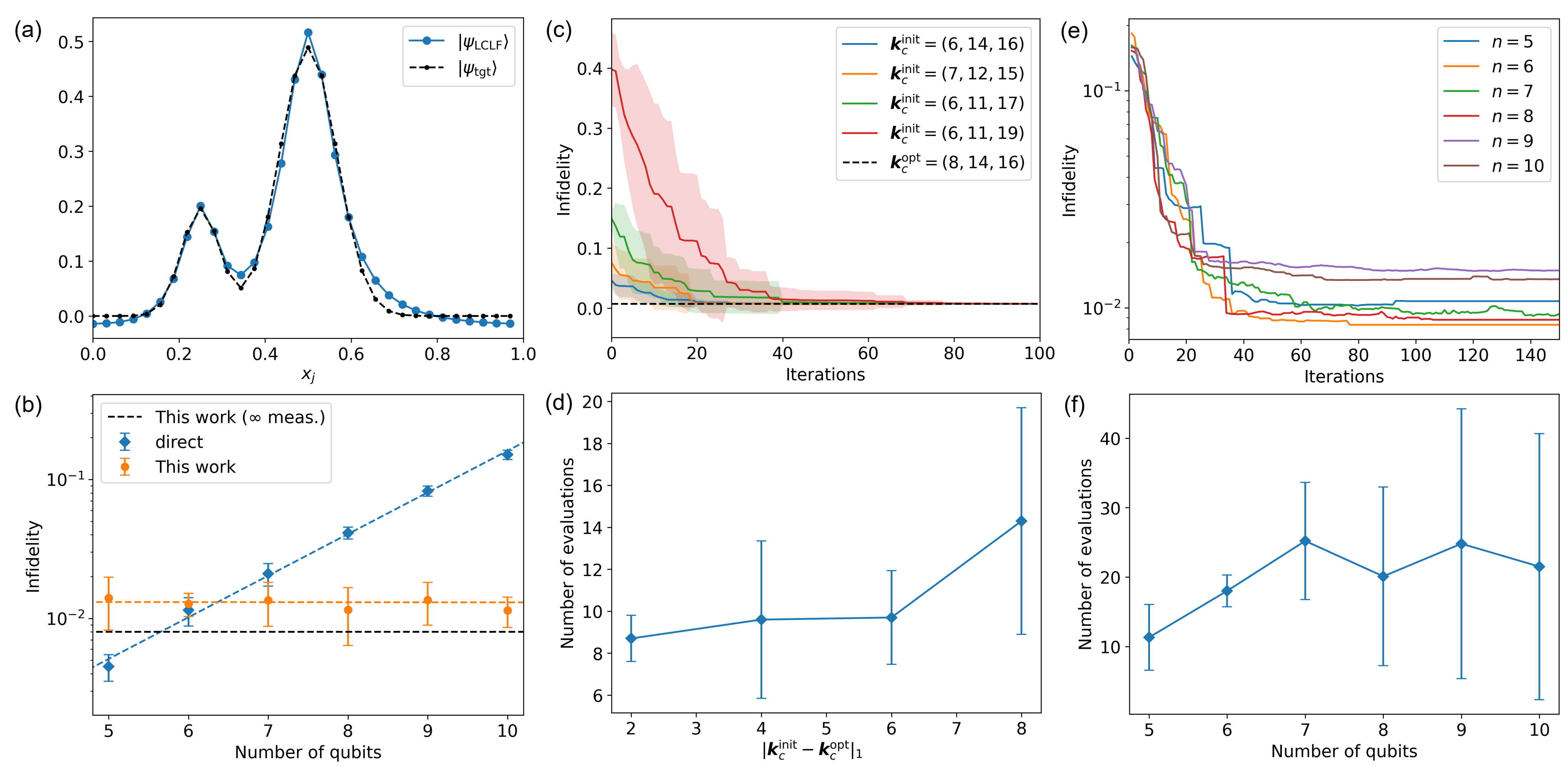} 
\caption{
(a) Target state $|\psi_{\text{tgt}}\rangle$ for $n=5$ qubits and the LCNOF state $|\psi_{\text{LCNOF}} (\boldsymbol{d}, \boldsymbol{a}, \boldsymbol{k}_c)\rangle$ ($n_{\text{bas}}=3$) that optimally approximates the target state. 
(b) Infidelity $\delta = 1 - |\langle \psi_{\text{tgt}}|\psi \rangle|^2$ according to the number of qubits, where $|\psi\rangle$ is the reconstructed state. 
Results are shown for direct measurement and for the proposed method, both using 1,000 measurements.
The black broken line represents the infidelity between $|\psi_{\text{tgt}}\rangle$ and $|\psi_{\text{LCNOF}}\rangle$ ignoring the statistical error.
Fidelity optimization using the Metropolis algorithm for various (c) initial peak centers and (e) number of qubits. Number of quantum overlap calculations as a function of (d) $\ell_1$ distance between the initial and optimized centers and  (f) number of qubits. $n=5$ was used in Figs. ~(c) and (d). $\boldsymbol{k}_c^{\mathrm{init}}/N = (3/16, 11/32, 17/32)$ are shown in Fig.~(e) and (f). In Figs.~(b–f), the results represent the average of ten independent trials. Standard deviations across these trials are shown as error bars or shaded regions.
}
\label{fig:result_readout_state}
\end{figure*}

In this study, we employed the same target state as in Ref.~\cite{Kosugi2024PRA} for our numerical simulations, given by
$
    \psi_{\text{ideal}}(x_j)
    \propto
    \exp (- (32 (x_j - 0.5) / 3)^2 )
    +
    0.4 \exp(- (16(x_j - 0.25))^2 )
$
. This quantum state is designed to model the simulation of a first-quantized Hamiltonian in quantum chemistry calculations \cite{Kassal2008PNAS, Kosugi2022PRR, Nishiya2024PRA}.
Figure~\ref{fig:result_readout_state}(a) presents both the target state $|\psi_{\text{tgt}}\rangle$ and the state approximated using a linear combination of three LF states $|\psi_{\text{LCNOF}}(\boldsymbol{d}, \boldsymbol{a}, \boldsymbol{k}_c)\rangle$. The decay rates, peak centers, and expansion coefficients of the three LF states were $\boldsymbol{a} = (0.360, 1.672, 0.490)$, $\boldsymbol{k}_c/N=(0.25, 0.4375, 0.5)$, and $\boldsymbol{d}=(0.380, -0.517, 1.272)$, respectively.
The infidelity between the target and the LCNOF state was $\delta = 7.1 \times 10^{-3}$.

Figure \ref{fig:result_readout_state}(b) presents a numerical calculation example for determining the expansion coefficient $\boldsymbol{d}$, given fixed peak centers $\boldsymbol{k}_c$ and decay rates $\boldsymbol{a}$.
We compared our approach with direct-measurement method, which is also referred to as the real-space readout (RSR) method, in contrast to the FSR method \cite{Huang2026QST}.
The infidelity of the direct measurement is fitted by $\delta = 0.15\times 2^n/N_{\mathrm{shot}}$, indicating that the infidelity increases exponentially with the number of qubits. It has been confirmed that an exponential number of measurements is required to accurately readout the quantum state via direct measurement as the number of qubits increases. 
In contrast, the infidelity associated with the reconstructed states obtained using the proposed method remains independent of the number of qubits. The results demonstrate that even for large quantum systems, the target state can be accurately reconstructed with a limited number of measurements.

Next, we examine the optimization of the peak positions. Here, the effects of statistical errors due to the finite number of measurements were not considered.
The inverse temperature in the Metropolis algorithm was varied with the number of steps as \cite{Okuyama2019PRE}
\begin{gather}
    \beta_k = \beta_0 \ln (1+k) .
\end{gather}
Additionally, the update magnitude for the peak positions was set to decrease as the optimization  proceeds:
\begin{gather}
    \Delta k
    =
    \max\left(
    \lceil 
        \alpha_0 - k/\alpha_1 
    \rceil, 1 
    \right).
\label{eq:update_rule_of_k}
\end{gather}

Figure~\ref{fig:result_readout_state}(c) shows the outcome of peak position optimization for four different sets of initial peak positions in the case of five qubits.
We used $\Delta k = 1$ and $\beta_0=100$.
It was observed that, on average, the number of iterations increases as the initial peak positions deviate further from the optimal ones. The number of iterations in this context indicates those involved in the classical Metropolis algorithm. The quantum overlap calculations for previously evaluated peak positions are omitted.
Consequently, the number of overlap calculations for unique peak positions emerges as a more relevant metric of the computational cost associated with quantum computations. This quantity is demonstrated in Fig.~\ref{fig:result_readout_state}(d). It was observed that the number of quantum overlap calculations increases approximately in proportion to the $\ell_1$ distance between the initial and optimal peak positions.

We investigate the dependence of the peak position optimization on the number of qubits for $\boldsymbol{k}_c^{\mathrm{init}}/N = (3/16, 11/32, 17/32)$. 
We used $\alpha_0 = 2^{n-5}$ and $\alpha_1=15$.
For $n=5$, the initial inverse temperature was set to $\beta_0 = 100$, while for $n \geq 6$, it was set to $\beta_0 = 150$. The convergence of the optimization, as measured by the decrease in infidelity, is shown in Fig.~\ref{fig:result_readout_state}(e). In the optimization process, the procedure was terminated once the infidelity dropped below 0.01.
For $n=9$ and $n=10$, the final infidelity remained above 0.01 in 1 out of 10 trials, in which the optimal peak positions were not found. The dependence of the number of quantum overlap evaluations on the number of qubits is shown in Fig.~\ref{fig:result_readout_state}(f). It was observed that the number of overlap evaluations does not significantly depend on the number of qubits.
In other words, this result indicates that the proposed method scales favorably to larger systems. This behavior may be attributed to the update rule of peak positions, defined as in Eq.~(\ref{eq:update_rule_of_k}). By employing more sophisticated strategies for determining the peak positions, such as a coarse-to-fine multilevel optimization strategy in which the peak positions are first optimized on a coarse grid and then successively refined on finer grids~\cite{Walshaw2004AOR, Alexander2009MathGeo, Pant2015PRE}, it may be possible to improve the stability of reaching the target infidelity while maintaining a feasible number of evaluations for larger numbers of qubits.
To sum up, the present readout method has the potential to outperform direct readout approaches as the number of qubits increases.

\subsection{Readout of QPE spectrum}
\begin{figure*}[ht]
    \centering
    \includegraphics[width=1.0 \textwidth]{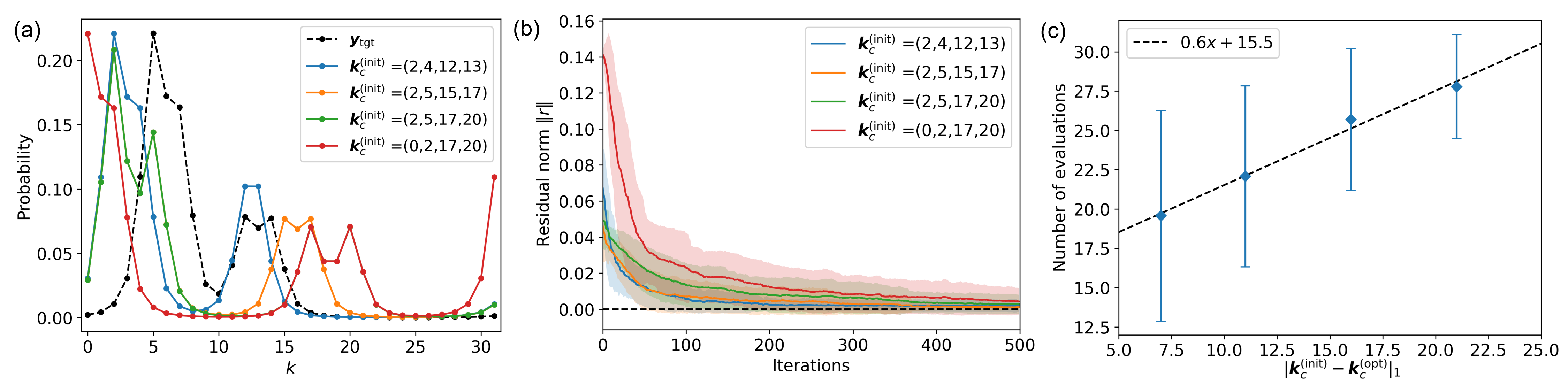} 
\caption{
(a) Probability distribution of target $\boldsymbol{y}_{\mathrm{tgt}}$ and initial probability distributions $\boldsymbol{y}_{\text{LCNOF}}(\boldsymbol{d}, \boldsymbol{a}, \boldsymbol{k}_c^{(\text{init})})$. (b) Residual norm according to number of iterations in the proposed method. (c) Number of quantum overlap evaluations $m_{\mathrm{eval}}$ on a quantum computer according to the $\ell_1$ distance between the target and initial peak centers as $\|\boldsymbol{k}_c^{(\mathrm{tgt})} - \boldsymbol{k}_c^{(\mathrm{init})}\|_{1}$. The shaded regions in Fig.~(b), and the error bars in Fig.~(c) represent the standard deviation of the results obtained from the 50 trials. In this figure, the broadening and initial inverse temperatures were set to $\eta =0.3$ and $\beta_0=100$, respectively.
}
\label{fig:result_ideal}
\end{figure*}

This section presents a calculation example that estimates the probability distribution of a quantum state using the method described in Sect.~\ref{sect:fit}. The probability distribution of the quantum state to be estimated $\boldsymbol{y}_{\mathrm{tgt}}$, as well as those of the initial peaks $\boldsymbol{y}_{\text{LCNOF}}(\boldsymbol{d}, \boldsymbol{a}, \boldsymbol{k}_c^{(\text{init})})$, are illustrated in Fig.~\ref{fig:result_ideal}(a).
In this example, the probability distribution of the target state is expressed as a linear combination of the square of LFs, and the peak widths are all equal and known in advance. In the calculation of XAS spectra using QPE sampling, the broadening parameter of the LF can be set as an input parameter, so this problem setting can be considered reasonable.

Figure~\ref{fig:result_ideal}(b) illustrates the relationship between the number of iterations $n_{\mathrm{iter}}$ and the residual norm in optimization for readout of quantum states. It has been demonstrated that the distinction between the initial peaks and the peaks of the target probability distribution is directly related to the magnitude of the residuals and the number of iterations necessary for convergence.

The number of iterations $n_{\mathrm{iter}}$ in the context is for classical computation. The number of quantum overlap calculations $m_{\mathrm{eval}}$ is a critical factor in comprehending the merits of this proposed measurement scheme. 
The results for the dependency of $m_{\mathrm{eval}}$ are presented in Fig.~\ref{fig:result_ideal}(c). Given the implementation of the Metropolis method, 50 trials were conducted, and the standard deviation is represented as error bars in Fig.~\ref{fig:result_ideal}(c). The horizontal axis of this figure denotes the $\ell_1$ distance between the peaks of the target quantum state and initial peaks. Numerical experimentation demonstrated a proportional relationship between the number of SWAP tests executed for the inner product calculation $m_{\mathrm{eval}}$ and the $\ell_1$ distance. In the initial state settings of this study, a mean of 28 overlap calculations was performed for the worst initial peaks. Considering the estimation of the coefficients using QAE for each computational basis function state, the number of overlap calculations is fewer than that of the bases, $2^{5}=32$, thereby demonstrating the superiority of the proposed method. Furthermore, in XAS spectra, the approximate peak positions can be estimated in advance using mean-field approximations \cite{Hohenberg1964, Kohn1965, Taillefumier2002PRB, Gao2008PRB, Gougoussis2009PRB}. Therefore, the number of quantum overlap calculations required for peak position optimization is expected to remain relatively small.

\section{Discussion}
\label{sec:discussions}
\subsection{Applicability to Various Quantum States}
\subsubsection{Number of Basis Functions in Larger Systems}
In Section  ~\ref{sect:numerical_results}, we consider a linear combination of one-dimensional Gaussian functions as a representative example of the proposed approach. We showed that as the grid size increases, the proposed method achieves lower infidelity than the RSR method for the same number of measurements. This improvement stems from the fact that the target quantum state can be efficiently represented by a linear combination of only a small number of LFs.
In Sect.~\ref{sect:numerical_results}, we restricted the approximation of the target function to a linear combination of three LFs. Consequently, the infidelity was limited to the order of $10^{-2}$. 
A smaller infidelity is expected to be attainable by increasing the number of basis functions.

In this section, we examine how many LFs are required to achieve a lower infidelity, in order to assess the applicability of the proposed method to more practical settings. We also studied the number of LFs required for two-dimensional target states. As an example of such a state, we consider a linear combination of two-dimensional Gaussian functions, defined in 
$
    \psi_{\mathrm{ideal}}(x_j,y_k)
    \propto
    \exp(-25((x_j-0.65)^2+(y_k-0.65)^2))
    +
    \exp(-16((x_j-0.35)^2+(y_k-0.35)^2))
$.
%
We define the required number of LFs, $n_{\mathrm{bas}}^{*}$, operationally as the first value of $n_{\mathrm{bas}}$ at which the infidelity reaches $10^{-4}$ in the following greedy construction. After $n_{\mathrm{bas}}$ LFs have been determined, the center of the ($n_{\mathrm{bas}}+1$)th LF is chosen from the set of maximizers of the fidelity. If multiple centers give the same maximum fidelity, we choose the first maximizer encountered in the numerical search. Throughout this procedure, the width $a$ of the LFs is fixed.

\label{sect:discussions}
\begin{figure}[ht]
    \centering
    \includegraphics[width=0.35 \textwidth]{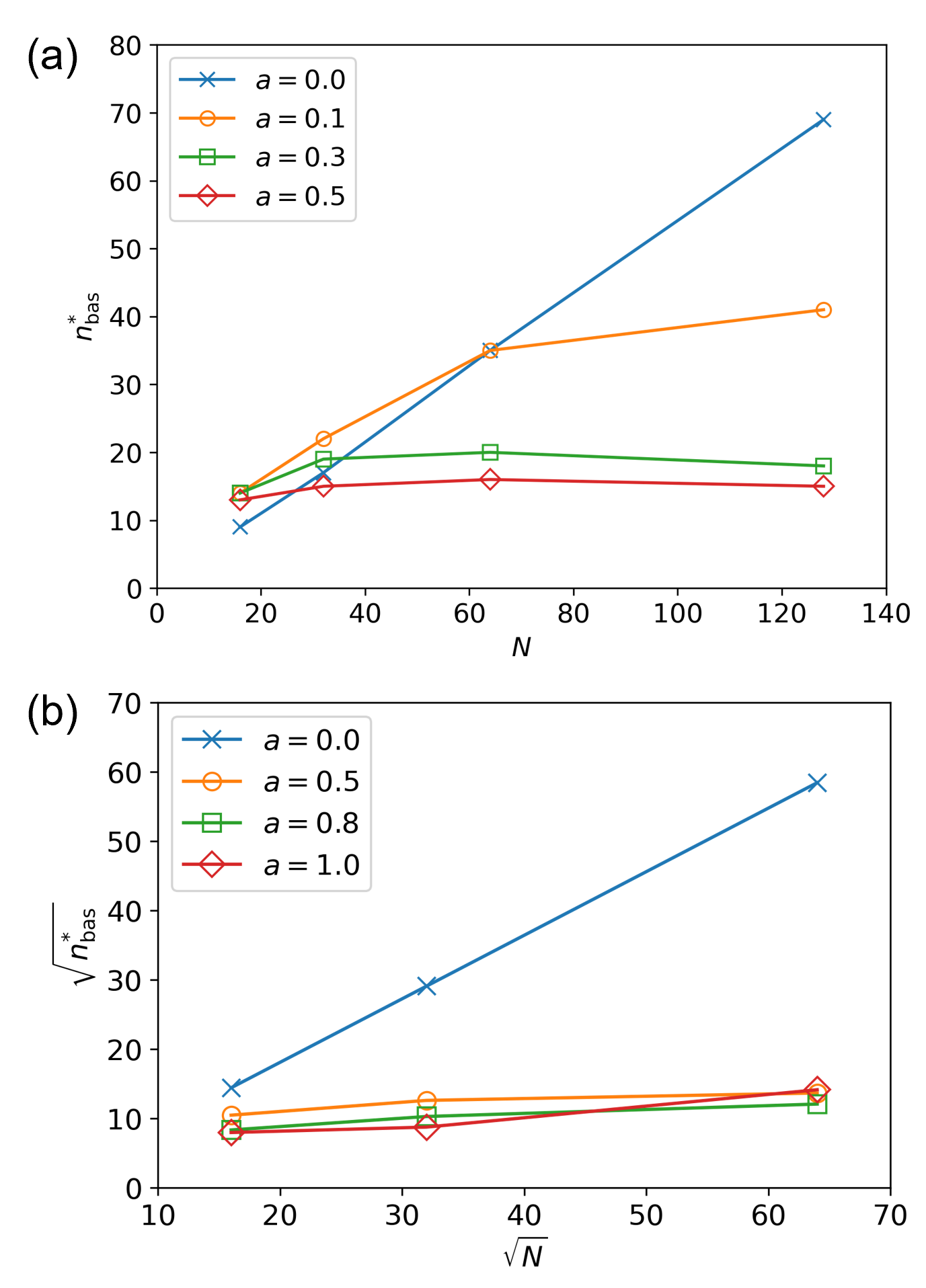}
\caption{The required number of LFs, $n_{\mathrm{bas}}^{*}$, to achieve an infidelity of $10^{-4}$ according to the grid size $N$ for target states represented by linear combinations of Gaussian functions: (a) one-dimensional and (b) two-dimensional states. }
\label{fig:nbas_opt}
\end{figure}

Figure~\ref{fig:nbas_opt} shows the dependence of the number of LFs $n_{\mathrm{bas}}^{*}$ required to achieve an infidelity of $10^{-4}$ on grid size $N$. For LFs with zero width, $a=0.0$, the required number of functions increases in proportion to the grid size. This behavior corresponds to the RSR method, in which the number of measurements increases with the grid size.
In contrast, when finite-width LFs are used, the required number of functions increases much more slowly with the grid size, and in some cases, it nearly saturates to a constant value. This result indicates that the number of LFs required to represent the target state has only a weak dependence on the grid size.
However, in more practical situations, the target quantum state may become more complex as the grid size increases, and a larger number of basis functions may be required. A systematic investigation of this aspect will be conducted in future studies.

\subsubsection{Generic highly entangled state}
It should be noted that the present framework does not enable efficient tomography of arbitrary quantum states. In particular, it is not efficient for generic, highly entangled states. Here, “generic” refers to states without special structures, such as permutation symmetry, sparsity, or an efficient tensor-network representation; for Haar-typical pure states, the bipartite entanglement entropy is known to be nearly maximal, indicating that such states are highly entangled in a generic sense. Our method assumes that each basis function can be prepared efficiently, and its advantage is obtained only when the target state can be represented as a linear combination of these basis functions. If each basis function has a bond dimension of at most $\chi_{\mathrm{max}}$, then a state represented as a linear combination of $n_{\mathrm{bas}}$ such basis functions has a bond dimension of at most $n_{\mathrm{bas}} \chi_{\mathrm{max}}$. Consequently, to represent a state with a Schmidt rank of order $\sqrt{N}$ across a balanced bipartition, the number of basis functions must satisfy $n_{\mathrm{bas}} \geq \sqrt{N} / \chi_{\mathrm{max}}$. 
While previous study has shown numerically that a one-dimensional Gaussian distribution can be represented with a small maximum bond dimension \cite{Rodriguez2024arXiv}, we have confirmed that a one-dimensional Lorentzian function exhibits a comparable maximum bond dimension.
This implies that the efficient reconstruction of generic highly entangled states is generally beyond the reach of the present scheme. More broadly, this limitation is consistent with the known exponential cost of general many-body quantum state tomography, whereas scalable tomography becomes possible only when additional structures are available, such as low-rank MPS structures.

\subsection{Relation to existing algorithms}
\subsubsection{Quantum state tomography and classical shadows}

QST has traditionally aimed at reconstructing a general density operator, applicable to both pure and mixed states,
\begin{gather}
    \rho 
    = 
    \sum_{i=1}^{r} p_i 
    |\psi_i\rangle\langle \psi_i|,
\end{gather}
where $|\psi_i\rangle$ is a normalized pure state, $p_i \ge 0$, and $\sum_{i=1}^{r} p_i = 1$. Here, $r$ denotes the rank of the mixed state. In this most general setting, QST seeks to determine all independent parameters of $\rho$ from the outcomes of informationally complete measurements. Early studies mainly focused on general reconstruction protocols, such as linear inversion and maximum likelihood approaches \cite{Hradil1997PRA, James2001PRA}, which do not assume any particular structure in the target state and are applicable to arbitrary pure or mixed states. Although these methods provide a broadly valid framework, the number of parameters to be estimated scales as $O(N^2)$ for a grid size $N$, and the required measurement and post-processing costs rapidly become prohibitive as the grid size increases.

To overcome this fundamental scalability issue, several studies have sought to reduce the measurement cost of QST by exploiting the structural properties of the target state. In particular, compressed sensing tomography by Gross \textit{et al.} showed that low-rank states, that is, states with small rank $r$, can be reconstructed from substantially fewer measurement settings than those required by generic $N^2$-scaling tomography \cite{Gross2010PRL}. Cramer \textit{et al.} further demonstrated that quantum states in one dimension that are well approximated by MPS can be reconstructed efficiently, including schemes based solely on local measurements \cite{Cramer2010Ncom}. Subsequent work has extended this line of research toward mixed-state tensor-network representations, such as matrix-product operators, as well as toward variational tensor-network tomography for more challenging higher-dimensional states \cite{Zhen2024IEEE, Guo2024CommPhys, Teng2025PRXQ}.


These studies primarily focus on quantum states arising in many-body systems. In contrast, recent interest in CAE/CFD has motivated readout and tomography protocols tailored to real-valued quantum states \cite{Manzano2023EPJ, Song2026CommPhys, Patterson2025QST}.  
By restricting the state space to real amplitudes, the cost of full QST can be reduced from $O(N^2)$ for general complex-valued quantum states to $O(N^{\log_2 3})$. Song \textit{et al.} proposed Hadamard Random Forest and numerically and experimentally demonstrated lower readout cost and higher fidelity than full QST on an IBM quantum processor \cite{Song2026CommPhys}.
Related readout strategies based on real-valued amplitude estimation have also been proposed \cite{Manzano2023EPJ, Patterson2025QST}. However, because these approaches typically do not assume further structure in the amplitudes, their classical post-processing remains at least linear in $N$, which can limit the practical benefit of the quantum speedup.

In this context, our work is motivated by the tomography of computational results produced by quantum simulations for applications such as CAE/CFD and XAS spectral calculations. Therefore, we focus on real-valued pure states while further exploiting the problem-specific structure in the target states to enable a more efficient readout.
The FSR method was proposed in the same spirit \cite{Huang2026QST, Huang2025arXiv2}.
Our method shares a similar idea by assuming that the real-valued quantum states can be expressed efficiently as linear combinations of a small number of non-orthogonal basis function states. We further require that these basis function states can be prepared efficiently, for example, through specific quantum gates or MPS encoding. Therefore, the method is expected to be effective for pure states with such low-complexity structures. 
A direct comparison with the approach of Cramer \textit{et al.} and the FSR method remains an interesting direction for future work.

Classical shadow tomography provides an efficient framework for estimating the expectation values of many observables, $\{\hat{\mathcal{O}}_i\}_{i=1}^{M}$, from a limited number of copies of an unknown quantum state. For pure states, Grier, Pashayan, and Schaeffer analyzed the task of predicting $\operatorname{Tr}(\hat{\mathcal O}\rho)$ for Hermitian observables satisfying $\|\hat{\mathcal O}\|_\infty \leq 1$ and $\operatorname{Tr}(\hat{\mathcal O}^2)\leq B$ \cite{Grier2024Quantum}. Applying their independent-measurement upper bound together with a union bound over $M$ observables, the sample complexity for a pure state in an $N$-dimensional quantum state is
\begin{gather}
    O\left(
        \min\left\{
            \frac{B}{\varepsilon^2},
            \frac{\sqrt{BN}}{\varepsilon}
            +
            \frac{1}{\varepsilon^2}
        \right\}
        \log \left(\frac{M}{\eta}\right)
    \right),
\label{eq:sample_complexity_shadow_pure_state}
\end{gather}
where $\eta$ denotes the target total failure probability and $\varepsilon$ is the target additive error in each observable estimate. For reconstructing real-valued pure states using Pauli expectation values, one needs to estimate $M=N^{\log_2 3}$ Pauli observables generated by tensor products of $X$, $Z$, and $I$. Each such Pauli observable satisfies $\|\hat{\mathcal O}\|_\infty=1$ and $\operatorname{Tr}(\hat{\mathcal O}^2)=N$, and hence $B=N$. For constant $\eta$, this gives the estimate
\begin{gather}
    O\left(
        \min\left\{
        \frac{N}{\varepsilon^2},
        \frac{N}{\varepsilon}
        +
        \frac{1}{\varepsilon^2}
    \right\}
    \log N
    \right).
\end{gather}
Thus, although classical shadows reduce the dependence on the number of observables from linear in $M$ to logarithmic in $M$, this estimate still scales linearly in $N$. In the joint-measurement setting, the pure-state classical-shadow bound improves the leading $B$-dependence to $\sqrt{B}$, but such protocols require coherent measurements across multiple copies and are therefore much more demanding experimentally \cite{Grier2024Quantum}. The above estimate should therefore be viewed as an order-of-magnitude comparison for applying classical shadows to real-valued pure-state reconstruction, rather than as a tomography-specialized optimal bound. We also note that real classical shadows have recently been developed for expectation-value estimation of real observables \cite{West2025JPAMT}, although a resource estimate specialized to the tomography of real-valued pure states is not yet available.

Our approach addresses a different objective. Rather than estimating many observables of an otherwise unstructured state, we seek to reduce the readout cost by exploiting the additional structure of the target pure state, such as localization or a compact representation in a tailored basis. Therefore, the two approaches are complementary: classical shadows are suited for predicting many observables, whereas our method is designed for efficient reconstruction under explicit structural assumptions. 
We also note that overlap and fidelity estimation tasks have already been investigated in the classical shadow framework, including recent shadow-overlap protocols for efficient state verification \cite{Struchalin2021PRXQ, Huang2025NatPhys, Li2025QIP}. Therefore, it would be interesting to investigate whether these ideas can be incorporated into our setting to further accelerate the evaluation of overlaps between the target state and the basis functions used in our reconstruction scheme.

\subsubsection{Ground state calculation algorithms}
We comment on the relationship between the present approach and ground-state calculation methods for many-body problems that employ multireference states. The present method may be viewed as a framework for maximizing the expectation value of the Hermitian operator
$
    \hat{\mathcal{O}}
    =
    |\psi_{\mathrm{tgt}}\rangle\langle \psi_{\mathrm{tgt}}|
$
with respect to the trial state
$
    |\psi_{\mathrm{LCNOF}}(\boldsymbol{d}, \boldsymbol{\theta})\rangle
$. 
Thus, the full quantum state is reconstructed. In contrast, ground-state calculation methods minimize the expectation value of the Hamiltonian,
$
    \hat{\mathcal{O}}=\hat{\mathcal{H}},
$
with respect to the many-body wave function. Various many-body wave-function ans\"atze and corresponding procedures for determining their coefficients have been proposed in the literature.

Quantum Krylov subspace expansion approaches \cite{Stair2020JCTC, Motta2020NPhys, Seki2021PRXQuantum} approximately construct the Krylov subspace
\begin{gather}
    \left\{
        |\phi\rangle,
        \hat{\mathcal{H}}|\phi\rangle,
        \hat{\mathcal{H}}^2|\phi\rangle,
        \cdots
    \right\}
\end{gather}
by means of real-time or imaginary-time evolution. Because of the approximation, the basis function states of the Krylov subspace are not exactly orthogonal to each other. Therefore, a generalized eigenvalue problem is constructed and solved. Unlike our approach, these methods do not introduce variational parameters into the reference states; however, they determine the coefficients of the linear combination by solving a generalized eigenvalue equation.

Multireference unitary coupled-cluster (MR-UCC) approaches \cite{Sugisaki2022PCCP, Guo2024Nphys, Wu2025arXiv} and subspace diagonalization approaches \cite{McClean2017PRA, Nakanishi2019PRR, Parrish2019PRL} share the common strategy of combining a multireference state with an additional shared parameterized unitary. A generic form of the ans\"atze may be written as
\begin{gather}
    |\psi\rangle
    =
    \hat{U}_{\mathrm{SP}}(\boldsymbol{\theta})
    |\psi_{\mathrm{ref}}\rangle
    =
    \hat{U}_{\mathrm{SP}}(\boldsymbol{\theta})
    \sum_{\ell=0}^{n_{\mathrm{bas}}-1} d_{\ell} |\phi_{\ell}\rangle,
\end{gather}
where \(\{|\phi_{\ell}\rangle\}\) is a set of orthogonal states satisfying \(\langle \phi_i | \phi_j \rangle = \delta_{ij}\), and the multireference state 
$
    |\psi_{\mathrm{ref}}\rangle 
    =
    \hat{U}_{\mathrm{MR}}(\boldsymbol{d})
    |0\rangle
    = 
    \sum_{\ell=0}^{n_{\mathrm{bas}}-1} d_{\ell} |\phi_{\ell}\rangle
$ 
is prepared using a parameterized unitary gate 
$
    \hat{U}_{\mathrm{MR}}(\boldsymbol{d})
$. 
The additional shared parameterized unitary \(\hat{U}_{\mathrm{SP}}(\boldsymbol{\theta})\), whose parameters are common to all reference states, acts on the multireference state to incorporate correlation effects.

In the MR-UCC approach, the UCC ansatz is used as $U_{\mathrm{SP}}(\theta) = e^{T(\boldsymbol{\theta}) - T^{\dagger}(\boldsymbol{\theta})}$ with the excitation operators $T(\boldsymbol{\theta})$. The multireference initial state can be prepared in several ways. Sugisaki \textit{et al.} used a classically precomputed multireference state \cite{Sugisaki2022PCCP}, whereas Guo \textit{et al.} \cite{Guo2024Nphys} used a unitary circuit with a single variational parameter, and Wu \textit{et al.} \cite{Wu2025PRA} employed a particle-number-conserving circuit \cite{Anselmetti2021NJP}. In the latter two approaches, the parameters $\boldsymbol{d}$ specifying the multireference initial state are first optimized by minimizing the energy expectation value within the VQE framework. After this optimization, these parameters were fixed, and only the UCC parameters {$\boldsymbol{\theta}$} were further optimized.

In contrast to the MR-UCC approach, the subspace diagonalization approach first optimizes the parameters of the shared parameterized unitary $U_{\mathrm{SP}}(\boldsymbol{\theta})$. The coefficients of the multireference state, $\{d_{\ell}\}$, are then determined by diagonalizing the subspace Hamiltonian on a classical computer.

In order to compare the MR-UCC, the subspace-diagonalization methods, and the proposed method on a common footing, we denote the basis states in the former methods as
$
    |f_{\ell}(\boldsymbol{\theta})\rangle 
    = 
    \hat{U}_{\mathrm{SP}}(\boldsymbol{\theta})|\phi_{\ell}\rangle
$,
where $\boldsymbol{\theta}$ is shared among all basis states.
The MR-UCC and subspace diagonalization methods differ from the proposed method in two respects: first, the parameters contained in the basis functions are common to all basis functions; second, the basis functions and their linear combination coefficients are not determined self-consistently in a one-shot procedure.

\section{CONCLUSIONS}
\label{sect:conclusions}
In this study, we proposed a method for quantum state readout and feature extraction using quantum-overlap-based fitting of function expansions. The methodology involves the calculation of the quantum overlaps between the non-orthogonal basis function state, such as discrete LF states, and the target state via measurements and classical optimization of the parameters in the function expansion. This approach is promising for scenarios in which the quantum state is expected to be represented as a linear combination of functional expansions. Furthermore, the proposed readout involves both the complex amplitudes and their squared magnitudes (probabilities) in the quantum state. The latter scenario arises in the calculation of spectral functions such as X-ray absorption spectra using quantum phase estimation. Preliminary numerical calculations demonstrated that this method reduced the number of measurements required to reconstruct a full pure quantum state.

The proposed approach can be extended to two- or three-dimensional systems. In such cases, the tensor decomposition technique will be useful for reducing the circuit depth to measure, as is the case with the amplitude-encoding approach \cite{Kosugi2025PRA}. The proposed approach utilizes the discrete LF state as the basis function state. However, another basis function state, that is the normal distribution \cite{Iaconis2024npjQI, Manabe2025Quantum}, is utilized for the reconstruction of full pure quantum states. The implementation of this study utilized a straightforward optimization procedure. Therefore, the implementation of a more sophisticated optimization algorithm will accelerate the process of obtaining the state to be estimated.

\section*{ACKNOWLEDGMENTS}
The authors acknowledge the contributions and discussions provided by the members of Quemix Inc.
The authors thank the Supercomputer Center, the Institute for Solid State Physics, the University of Tokyo for the use of the facilities.
This work was partially supported by the Center of Innovations for Sustainable Quantum AI (JST Grant Number JPMJPF2221).

\appendix
\section{Case where the numbers of qubits differ in the SWAP test}
\label{sect:overlap_diff_qubits}

Here, we consider the case where the numbers of qubits of the target
state and the basis-function state are different.

First, suppose that the basis-function state has $n+m$ qubits,
whereas the target state has $n$ qubits.
An index of the $(n+m)$-qubit basis-function state can be written as
$2^m k+j$, where
$k=0,\ldots,2^n-1$ labels the upper $n$ bits and
$j=0,\ldots,2^m-1$ labels the lower $m$ bits.
By summing over the lower $m$ bits, we define
\begin{gather}
    \bar{y}_{k}(\boldsymbol{\theta}_{\ell})
    \equiv
    \sum_{j=0}^{2^m-1}
    y_{2^m k+j}(\boldsymbol{\theta}_{\ell})
    =
    \sum_{j=0}^{2^m-1}
    \left|
        f_{2^m k+j}(\boldsymbol{\theta}_{\ell})
    \right|^2 .
\label{eq:marginal_basis}
\end{gather}
Then,
\begin{align}
    h_{\ell}(\boldsymbol{\theta})
    =
    \sum_{k=0}^{2^n-1}
    |c_k|^2
    \bar{y}_{k}(\boldsymbol{\theta}_{\ell})
    =
        \boldsymbol{y}_{\mathrm{tgt}}
        \cdot
        \bar{\boldsymbol{y}}(\boldsymbol{\theta}_{\ell})
    .
\label{eq:unequal_qubits_basis}
\end{align}

Conversely, suppose that the target state has $n+m$ qubits,
whereas the basis-function state has $n$ qubits.
Using the same indexing, $k$ labels the upper $n$ bits and $j$
labels the lower $m$ bits of the target state.
The marginalized target distribution is defined as
\begin{gather}
    \bar{y}_{\mathrm{tgt},k}
    \equiv
    \sum_{j=0}^{2^m-1}
    |c_{2^m k+j}|^2 .
\label{eq:marginal_target}
\end{gather}
Therefore,
\begin{align}
    h_{\ell}(\boldsymbol{\theta})
    =
    \sum_{k=0}^{2^n-1}
    \bar{y}_{\mathrm{tgt},k}
    y_k(\boldsymbol{\theta}_{\ell})
    =
        \bar{\boldsymbol{y}}_{\mathrm{tgt}}
        \cdot
        \boldsymbol{y}(\boldsymbol{\theta}_{\ell})
    .
\label{eq:unequal_qubits_target}
\end{align}

\section{Evaluation of error bound on solving the generalized eigenvalue equation}
\label{sect:error_bound}
In this section, we consider the effect of the statistical error when it appears in the evaluation of the quantum overlap $\bm{b}$ in the construction of the generalized eigenvalue equation in Eq.~(\ref{ampl_encoding_of_GMO:eig_prob_for_lc_coeffs}).
We simply write $\bm{b}(\bm{\theta}) = \bm{b}$.
We restrict the analysis to the case in which
$|\psi_{\mathrm{tgt}}\rangle$ and the quantum overlaps $b_\ell$ are real-valued. 
Let $\hat{e}_{\ell}$ denote a random variable describing the statistical error that occurs when evaluating  $b_{\ell}$. 
The random variable satisfies
\begin{gather}
\begin{aligned}
    \mathbb{E}[\hat{e}_{\ell}] &= 0 
    \\
    \mathrm{Var}[\hat{e}_{\ell}]
    &=
    \frac{
        1 - |b_{\ell}|^2
    }{N_{\mathrm{shot}}},
\label{eq:def_random_variables}
\end{aligned}
\end{gather}
where $N_{\mathrm{shot}}$ is the number of measurements.
In the first order of $e$, the matrix $G$ in the generalized eigenvalue equation is changed:
\begin{gather}
    G
    \to
    G
    +
    \mathcal{E}
    ,
\end{gather}
where
\begin{gather}
    \mathcal{E}_{\ell, \ell^{'}}
    \equiv
    \hat{e}_{\ell^{\prime}}
    b_{\ell}
    + 
    \hat{e}_{\ell}
    b_{\ell^{'}}
\end{gather}
According to the first-order perturbation theory, the eigenvalue and eigenstate corrections of the generalized eigenvalue equation can be written as \cite{JJSakurai}
\begin{gather}
\begin{aligned}
    \delta \lambda_{0} 
    &= 
    \bm{d}_0 \cdot \mathcal{E} \bm{d}_0.
    \\
    \delta \bm{d}_0
    & =
    \sum_{i>0}
    \frac{
        \bm{d}_{i}
        \cdot
        \mathcal{E} \bm{d}_0
    }{
        \lambda_{0}
        -
        \lambda_{i}
    }
    \bm{d}_i,
\end{aligned}
\end{gather}
where $\lambda_i$, $i=0,1,\ldots, n_{\mathrm{bas}}-1$, denote the eigenvalues sorted in descending order.
Then, the correction to the reconstructed state is given by 
\begin{gather}
    |\delta \psi\rangle
    =
    \sum_{\ell=0}^{n_{\mathrm{bas}}-1}
    \sum_{i > 0}
    \frac{
        \bm{d}_i \cdot \mathcal{E} \bm{d}_0
    }{
        \lambda_0 - \lambda_i
    }
    d_{i, \ell} |f_{\ell}\rangle .
\end{gather}
The infidelity due to the statistical error is approximately quantified using the norm of $|\delta \psi\rangle$:
\begin{gather}
\begin{aligned}
    \delta
    &=
    1 - |\langle\psi|\widetilde{\psi}\rangle|^2
    \leq
    \langle \delta \psi| \delta\psi\rangle
   \\&=
    \sum_{i>0}
        \frac{
        |\bm{d}_i \cdot \mathcal{E} \bm{d}_0|^2
    }{
        (\lambda_0 - \lambda_i)^2
    }
    \leq
    \Delta \lambda^{-2}
    \sum_{i>0}
    |\bm{d}_i \cdot \mathcal{E} \bm{d}_0|^2
    ,
\end{aligned}
\label{eq:infidelity_statistical_error}
\end{gather}
where $\Delta\lambda = \lambda_{0} - \lambda_{1}$ and $|\widetilde{\psi}\rangle$ denotes the normalized perturbation state in the first order, expressed as
\begin{gather}
    |\widetilde{\psi}\rangle
    =
    \frac{1}{
        \sqrt{1 + \langle \delta\psi|\delta\psi\rangle}
    }
    (|\psi\rangle+|\delta\psi\rangle)
    .
\label{eq:perturbation_state}
\end{gather}
The assumptions $\lambda_0 \approx 1$ and $\lambda_i \approx 0$ for $i>0$ are justified by the small statistical error.
In such a situation, $\Delta \lambda \approx 1$.
We have
\begin{gather}
    \sum_{i>0}
    |\bm{d}_i \cdot \mathcal{E} \bm{d}_0|^2
    =
    \bar{\bm{d}}_{0}\cdot \bar{\mathcal{E}}^2
    \bar{\bm{d}}_{0} 
    -
    (\bar{\bm{d}}_{0} \cdot \bar{\mathcal{E}} \bar{\bm{d}}_{0} )^2
\label{eq:pt_numerator}
\end{gather}
where $\bar{\bm{d}}_0 \equiv S^{\frac{1}{2}} \bm{d}_0$ and $\bar{\mathcal{E}} \equiv S^{-\frac{1}{2}} \mathcal{E} S^{-\frac{1}{2}}$.

Equation (\ref{eq:pt_numerator}) is bounded by \cite{Horn2012book}
\begin{gather}
    \bar{\bm{d}}_{0}\cdot \bar{\mathcal{E}}^2
    \bar{\bm{d}}_{0} 
    -
    (\bar{\bm{d}}_{0} \cdot \bar{\mathcal{E}} \bar{\bm{d}}_{0} )^2
    \leq
    \|\bar{\mathcal{E}}^2\|_2
    \leq
    \frac{
        \|\mathcal{E}\|_2^2
    }{\sigma_{\min}(S)^2},
\label{eq:pt_numerator_upper_bound}
\end{gather}
where $\sigma_{\min}(S)$ denotes the minimum singular value of $S$.
The variance of the statistical variables in Eq.~(\ref{eq:def_random_variables}) is bounded as
\begin{gather}
    \mathrm{Var}[\hat{e}_{\ell}]
    \leq
    \frac{1}{N_{\mathrm{shot}}} .
\label{eq:max_random_variance}
\end{gather}
Using Eq.~(\ref{eq:max_random_variance}), the perturbation matrix $\mathcal{E}$ satisfies the variance-based bound as
\begin{gather}
    \mathcal{E}
    \leq
    \frac{1}{\sqrt{N_{\mathrm{shot}}}}
    (
        \bm{b} \bm{1}^{T} 
        +
        \bm{1} \bm{b}^{T} 
    ),
\label{eq:pert_mat_upper_bound}
\end{gather}
where $\bm{1} \equiv (1,1,\ldots,1)$.
The matrix norm of $\mathcal{E}$ is evaluated as \cite{Horn2012book}
\begin{gather}
\begin{aligned}
    \|\mathcal{E}\|_2
    &\leq
    \frac{1}{\sqrt{N_{\mathrm{shot}}}}
    \|\bm{b} \bm{1}^{T} 
    +
    \bm{1} \bm{b}^{T} \|_{2}
    \\
    &=
    \frac{1}{\sqrt{N_{\mathrm{shot}}}}
    \left(
    \left|
        \sum_{\ell} b_{\ell}
    \right|
    +
    \sqrt{n_{\mathrm{bas}}}
    \|\bm{b} \|_2
    \right)
    \\ &\leq
    2\sqrt{
        \frac{n_{\mathrm{bas}}}{N_{\mathrm{shot}}}
    }
    \|\bm{b} \|_2 
    =
    2\sqrt{
        \frac{n_{\mathrm{bas}}}{N_{\mathrm{shot}}}
    }
    \sqrt{\bm{d}_0 \cdot S^2 \bm{d}_0}
    .
\label{eq:l2_norm_pt_matrix}
\end{aligned}
\end{gather}
Here, since we assume $|\psi_{\mathrm{tgt}}\rangle = \sum_{\ell}d_{0,\ell}|f_{\ell}\rangle$, we have $b_{\ell} = \sum_{m} S_{\ell,m} d_{0,m}$.
To sum up Eqs.~(\ref{eq:infidelity_statistical_error}), (\ref{eq:pt_numerator_upper_bound}), and (\ref{eq:l2_norm_pt_matrix}), we have
\begin{gather}
\begin{aligned}
    \delta 
    &\leq
    \frac{
        4 n_{\mathrm{bas}}\bm{d}_0 \cdot S^2 \bm{d}_0
    }{
        \Delta \lambda^{2}N_{\mathrm{shot}} \sigma_{\min}(S)^2
    }
    \\
    &\leq
    \frac{
        4 n_{\mathrm{bas}} \sigma_{\max}(S)
    }{
        \Delta \lambda^{2}N_{\mathrm{shot}} \sigma_{\min}(S)^2
    }   
    \\ &\leq
    \frac{
        4 n_{\mathrm{bas}} \kappa(S)^2
    }{
        \Delta \lambda^{2}N_{\mathrm{shot}} 
    }
    ,
\end{aligned}
\label{eq:infidelity_upper}
\end{gather}
where $
    \bm{d}_0 \cdot S^2 \bm{d}_0
    =
    \bar{\bm{d}}_{0}\cdot S \bar{\bm{d}}_{0} \leq \sigma_{\max}(S)
$ is used.
$\kappa(S) \equiv \sigma_{\max}(S)/\sigma_{\min}(S)$ denotes the condition number of $S$, and the last inequality follows from the fact that $S$ is Hermitian and positive semidefinite, with $\mathrm{tr}(S) = n_{\mathrm{bas}}$.
In the following, we numerically evaluate this inequality, which does not impose special conditions for the target state. However, the resulting upper bound is overly conservative because the estimate 
$
    \|\mathcal{E} S^{-1} \mathcal{E}\|_{2}
    \leq
    \|\mathcal{E}\|_{2}^{2} / \sigma_{\min}(S)
$
is not tight.

To simplify the discussion, we consider the situation in which the target states have equal linear-combination weights and the overlap matrix is Toeplitz.
This corresponds, for example, to equally spaced Lorentzian functions with the same broadening parameters.
In this case, the target state is written as
\begin{gather}
    |\psi_{\mathrm{tgt}}\rangle
    =
    \frac{1}{\sqrt{\sum_{i,j}S_{i,j}}}
    \sum_{\ell=0}^{n_{\mathrm{bas}}-1}
    |f_{\ell}\rangle,
\label{eq:target_state_uniform}
\end{gather}
which corresponds to
$
    \bm{d}_{0}
    =
    \bm{1} / \sqrt{\sum_{i,j}S_{i,j}}
$.
We evaluate an upper bound of Eq.~(\ref{eq:pt_numerator}) on this linear-combination coefficient as 
\begin{gather}
\begin{aligned}
    &
    \frac{1}{\sum_{i,j}S_{i,j}}
    \sum_{p,q}(\mathcal{E} S^{-1} \mathcal{E})_{p,q}
    -
    \left(
        \frac{1}{\sum_{i,j}S_{i,j}}
        \sum_{p,q}\mathcal{E}_{p,q}
    \right)^2
    \\ \leq &  
    \frac{1}{\sum_{i,j}S_{i,j}}
    \sum_{p,q}(\mathcal{E} S^{-1} \mathcal{E})_{p,q}
    \\ \leq & 
    \frac{n_{\mathrm{bas}}}{\sum_{i,j}S_{i,j}}
    \|\mathcal{E} S^{-1} \mathcal{E}\|_2
    \\ \leq &
    \frac{n_{\mathrm{bas}}}{\sum_{i,j}S_{i,j}}
    \frac{\|\mathcal{E}\|^2_2}{
        \sigma_{\min}(S)
    }
\label{eq:pt_numerator_upper_bound_v2}
\end{aligned}
\end{gather}
Also, in case of Eq.~(\ref{eq:target_state_uniform}) and $S$ is Toeplitz matrix, the right-hand-side of the first equation in Eq~(\ref{eq:l2_norm_pt_matrix}) is calculated as
\begin{gather}
    \|\bm{b} \bm{1}^{T} 
    +
    \bm{1} \bm{b}^{T} \|_{2}
    =
    2 \sqrt{\sum_{i,j} S_{i,j}}.
\label{eq:equaiton_overlaps_norm}
\end{gather}
Finally, to sum up Eqs.~(\ref{eq:infidelity_statistical_error}), 
(\ref{eq:l2_norm_pt_matrix}), 
(\ref{eq:pt_numerator_upper_bound_v2}), and (\ref{eq:equaiton_overlaps_norm}), we obtained the variance-based upper bound as
\begin{gather}
    \delta 
    \leq
    \frac{
        4 n_{\mathrm{bas}}
    }{
        \Delta \lambda^{2}N_{\mathrm{shot}}\sigma_{\min}(S)
    }
    \leq
    \frac{
        4 n_{\mathrm{bas}} \kappa(S)
    }{
        \Delta \lambda^{2}N_{\mathrm{shot}}
    }
    .
\label{eq:infidelity_upper_v2}
\end{gather}
These upper bounds in Eqs.~(\ref{eq:infidelity_upper}) and (\ref{eq:infidelity_upper_v2}) imply that the proposed readout method does not depend on the number of qubits, but on the number of basis functions $n_{\mathrm{bas}}$ and the orthogonality of the overlap matrix $S$ characterized by the condition number of $S$. 

\begin{figure*}[ht]
    \centering
    \includegraphics[width=0.9 \textwidth]{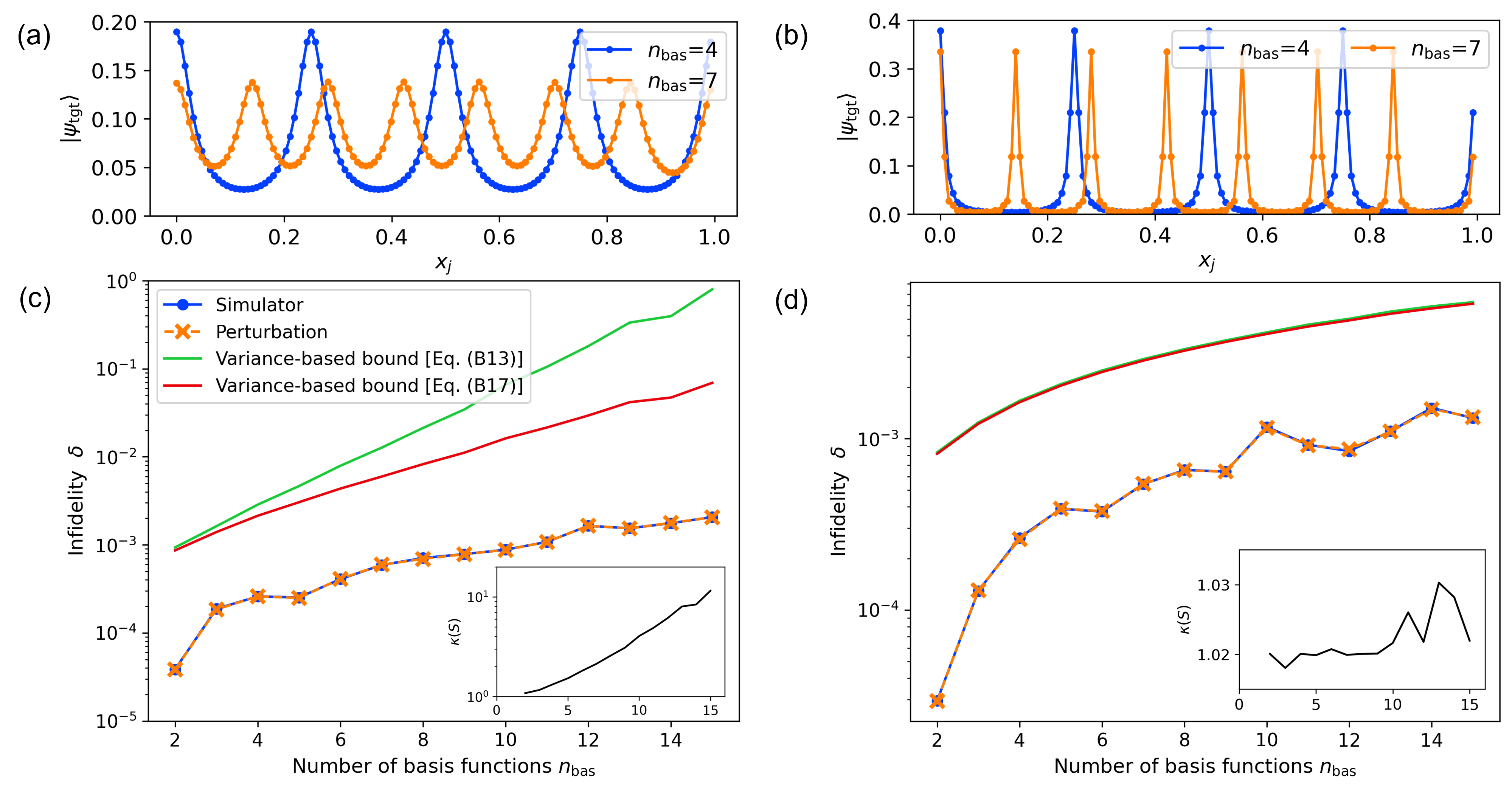}
\caption{
(a, b) Target state $|\psi_{\mathrm{tgt}}\rangle$ for $n=7$ qubits, with the Lorentzian broadening parameters set to (a) $a=0.2$ and (b) $a=0.2/n_{\mathrm{bas}}$.
(c, d) Infidelity between the target and reconstructed states as a function of the number of basis functions, for $N_{\mathrm{shot}}=10^{4}$. The infidelities are averaged over 10 samples. “Perturbation’’ denotes the approximation obtained from the first-order perturbation theory in Eq.~(\ref{eq:perturbation_state}). 
The variance-based bounds obtained from the first-order perturbative analysis, Eqs.~(\ref{eq:infidelity_upper}) and (\ref{eq:infidelity_upper_v2}), are shown for comparison, where $\Delta \lambda =1$ is used.
The inset shows the condition number $\kappa(S)$ of the overlap matrix as a function of $n_{\mathrm{bas}}$.
}
\label{fig:error_bound}
\end{figure*}

Figure \ref{fig:error_bound} shows the infidelity of the reconstructed states evaluated from the statistical errors. The upper bounds given by Eqs.~(\ref{eq:infidelity_upper}) and (\ref{eq:infidelity_upper_v2}) are also plotted for comparison. The target state $|\psi_{\mathrm{tgt}}\rangle$ is taken to be a linear combination of Lorentzian functions, where these peaks have the same coefficients and broadening parameters, and the peaks are arranged almost uniformly (exactly uniformly when divisible by the number of basis functions).

We consider two scenarios. First, all Lorentzian functions share the same broadening parameter, $a=0.2$, independent  of $n_{\mathrm{bas}}$, as shown in Fig.~\ref{fig:error_bound}(a). Second, the broadening parameter is scaled as $a=0.2/n_{\mathrm{bas}}$, as plotted in Fig.~\ref{fig:error_bound}(b).

The numerically obtained infidelities are well reproduced by the first-order perturbative theory, and the upper bounds correctly capture their dependence on the number of basis functions, as shown in Fig.~\ref{fig:error_bound}(c) and (d).
In Fig.~\ref{fig:error_bound}(c), the infidelity increases exponentially with increasing $n_{\mathrm{bas}}$, whereas it increases linearly in Fig.~\ref{fig:error_bound}(d).

The exponential growth of the infidelity is primarily caused by the exponential increase in the condition number $\kappa(S)$ of the overlap matrix, appearing in Eqs.~(\ref{eq:infidelity_upper}) and (\ref{eq:infidelity_upper_v2}), whose dependence is shown in the inset in Fig.~\ref{fig:error_bound}(c). 
The exponential increase of $\kappa(S)$ originates from the exponential decrease of the smallest singular value. 
By contrast, in Fig.~\ref{fig:error_bound}(d), the infidelity increases linearly with $n_{\mathrm{bas}}$, since $\kappa(S)$ is nearly independent of $n_{\mathrm{bas}}$.
The worsening of the condition number is attributed to the large overlap among the basis functions.
Therefore, it is necessary to choose basis function states that are as orthogonal as possible to each other. Moreover, the upper bound in Eq.~(\ref{eq:infidelity_upper_v2}) is tighter than that from Eq.~(\ref{eq:infidelity_upper}). This is because the inequality becomes tighter as the overlap matrix approaches the identity. It is also confirmed that the upper bound is tighter for smaller values of $\sigma_{\min}(S)$.

\section{Implementation of shifted discrete Lorentzian function state}
\label{sect:impl_discrete_lorentzian}
The shifted discrete LF state is written as  
\begin{gather}
\begin{aligned}
    |L; a, k_c\rangle
    &=
    T(k_c)|L; a, k_c =0\rangle
    \\ &=
    U_{\mathrm{QFT}}
    U_{\mathrm{shift}}(k_c)
    |S, a\rangle
    \\ &=
    U_{\mathrm{QFT}}^{\dagger}
    U_{\mathrm{shift}}(-k_c)
    |S, a\rangle,
\label{eq:c1}
\end{aligned}
\end{gather}
where $a > 0$ and $k_c$ denote the decay rate and the peak center of LF, respectively.
The inverse QFT in the translation operator cancels the QFT used to prepare the origin-centered LF state, yielding the first representation in Eq.~(\ref{eq:c1}). The equivalent second representation is used in Fig.~\ref{circuit:shifted_discrete_lf_state}.
Accordingly, the phase gates in Fig.~\ref{circuit:shifted_discrete_lf_state} are applied with angles $-\varphi_{m}$ defined in the following description.
The state $|S, a\rangle$ represents the discrete SF state with the decay rate $a$ where the center of the peak is origin.
The discrete SF state is generated using the following gate operations as
\begin{gather}
    |S, a \rangle 
    =
    \mathrm{C}X^{\otimes n -1}
    \bigotimes_{m=0}^{n-1}
    R_Y(2\theta_m)
    |0\rangle_{n}
\end{gather}
where
\begin{gather}
    \theta_m
    \equiv
    \begin{cases}
        \arctan e^{-a} & m=n -1 \\
        \arctan \exp(-2^m a) & \text{otherwise}
    \end{cases} ,
\end{gather}
and $\mathrm{C}X^{\otimes n -1}$ is known as the quantum fanout gate \cite{Fang2003arXiv, Takahashi2021TCS}.
The phase shift unitary is defined as
\begin{gather}
    U_{\mathrm{shift}}(k) |j\rangle_{n}
    =
    \exp\left(
        -i \frac{2\pi k}{N} j
    \right)
    |j\rangle_{n}
\end{gather}
and is implemented with the single-qubit phase shift gates $Z(\varphi)$  as
\begin{gather}
    U_{\mathrm{shift}}(k_c)
    =
    \bigotimes_{m=0}^{n-1}
    Z(\varphi_m),
\end{gather}
with the angle parameter
$
    \varphi_m
    =
    -2\pi k_c 2^m / N
$.

\begin{figure}[h]
\centering
\mbox{ 
\Qcircuit @C=1em @R=1.2em{
\lstick{|0\rangle} & \gate{R_Y(2\theta_0)} & \targ & \gate{Z({-\varphi_0})} & \multigate{2}{\text{QFT}^{\dagger}} & \qw \\
\lstick{|0\rangle} & \gate{R_Y(2\theta_1)} & \targ    & \gate{Z({-\varphi_1})} & \ghost{\text{QFT}^{\dagger}} &  \qw \\
\lstick{|0\rangle} & \gate{R_Y(2\theta_2)} & \ctrl{-2}    & \gate{Z({-\varphi_2})} & \ghost{\text{QFT}^{\dagger}} &  \qw \\
}
} 
\caption{
Quantum circuit for shifted discrete LF state in Eq.~(\ref{eq:c1}) for $n=3$.
}
\label{circuit:shifted_discrete_lf_state}
\end{figure}

\section{Dependency of overlap between Lorentzian function states}
\label{sect:appendix_overlap}

\begin{figure}[ht]
  \begin{minipage}[b]{0.85\hsize}
    \centering
    \includegraphics[width=0.85 \textwidth]{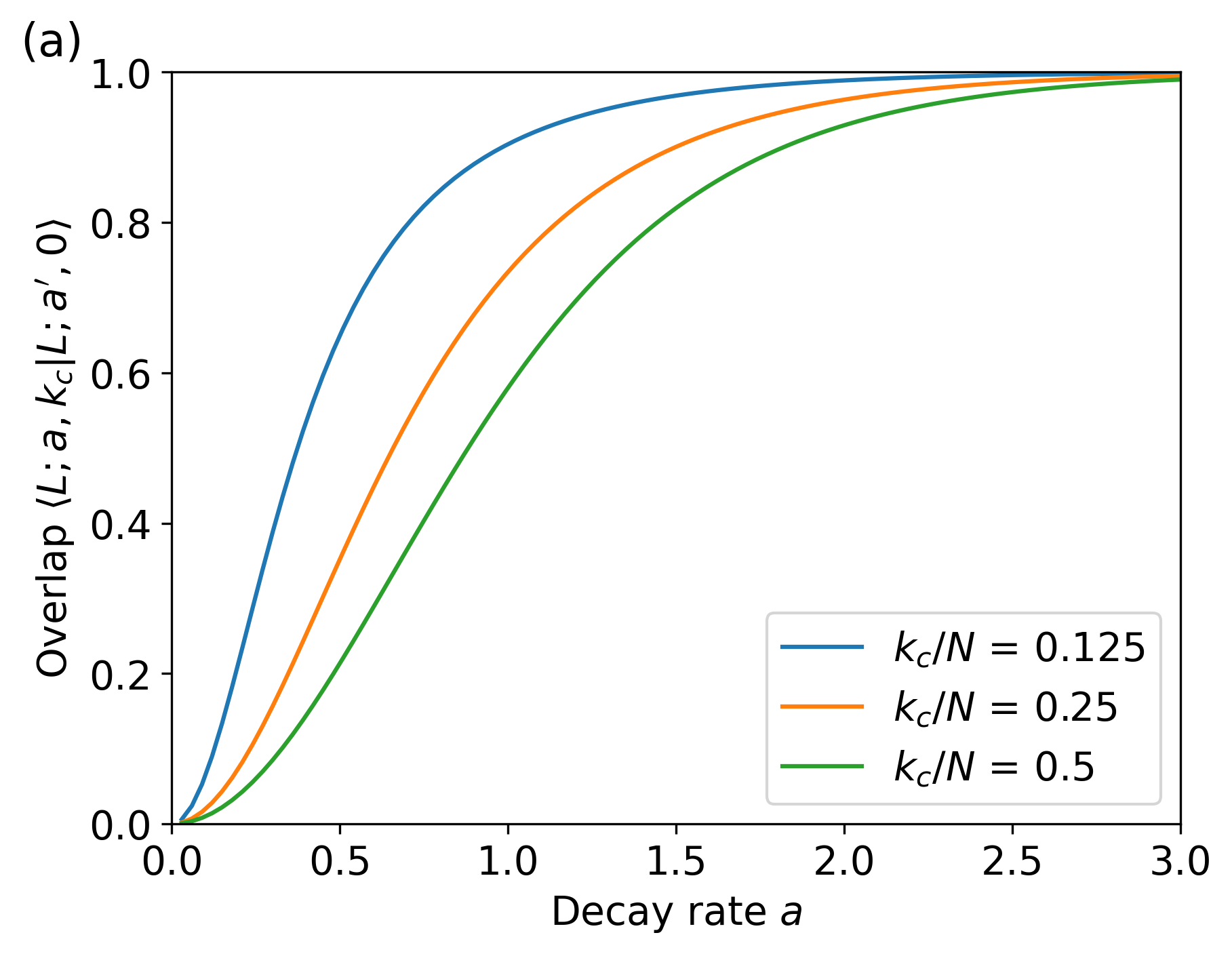}
  \end{minipage}
  \begin{minipage}[b]{0.85\hsize}
    \centering
    \includegraphics[width=0.85 \textwidth]{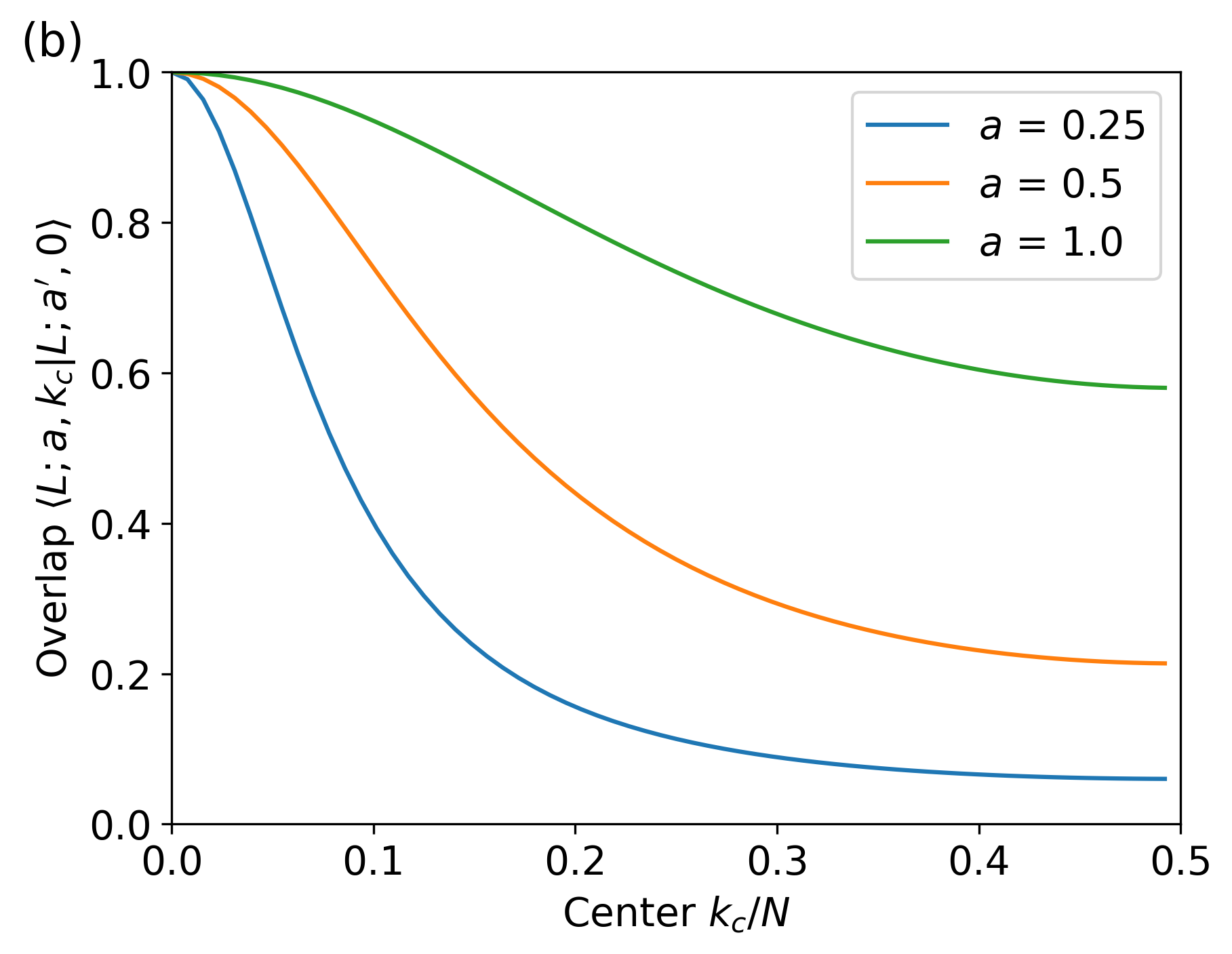}
  \end{minipage}
\caption{The overlap between two LFs in Eq.~(\ref{eq:overlap_two_lfs}) according to (a) the decay rate $a$ and (b) peak center $k_c$. We used $n=7$ and $a=a^{'}$ in this figure.}
\label{fig:overlap_two_lfs}
\end{figure}

The overlap between two LFs is given by
\begin{gather}
    \langle L; a, k_c | L; a^{\prime}, 0\rangle
    \notag \\
    =
    C_S(n, a) C_S(n, a^{\prime})
    \frac{
        (1 - (-1)^{k_c}e^{-\frac{(a+a^{\prime})N}{2}})
        \sinh{(a+a^{\prime})}
    }{
        \cosh{(a+a^{\prime})}
        -
        \cos{(2\pi k_c / N)}
    }
\label{eq:overlap_two_lfs}
\end{gather}
The derivation of the overlap is described in Appendix G in \cite{Kosugi2024PRA}.
For a large number of qubits, the overlap is independent on the number of qubits $n$.
The dependency of the overlap on the decay rate $a$ and the peak center $k_c$ is plotted in Fig~\ref{fig:overlap_two_lfs}.
As the decay rate increases, the distribution of LF states spreads and the overlap approaches 1. It has been confirmed that the overlap decreases monotonically as the peak positions of the two LF states move apart.

Also, the overlap between two squared of LFs is calculated as
\begin{gather}
    \sum_{k=0}^{N-1}
    L_{k}^2 (n, a)
    L_{k- k_{c\ell}}^{2} (n, a^{\prime})
    \notag \\
    =
    \frac{
        C_S^2(n, a) C_S^2(n, a^{\prime})
    }{N^2}
    \sinh^2{a} \sinh^2{a^{\prime}} 
    \notag \\
    \sum_{k=0}^{N-1}
    \frac{
        (1-(-1)^{k}e^{-aN/2})^2
        (1-(-1)^{k^{\prime}}e^{-a^{\prime}N/2})^2
    }{
        \left(
            \cosh{a}
            -
            \cos{\left(\frac{2\pi k}{N}\right)}
        \right)^2
        \left(
            \cosh{a^{\prime}}
            -
            \cos{\left(\frac{2\pi k^{\prime}}{N}\right)}
        \right)^2
    }
\label{eq:overlap2_two_lfs}
\end{gather}
where $k^{\prime} \equiv (k - k_{c\ell})~\mathrm{mod}~N$.

\begin{figure}[ht]
  \begin{minipage}[b]{0.85\hsize}
    \centering
    \includegraphics[width=0.85 \textwidth]{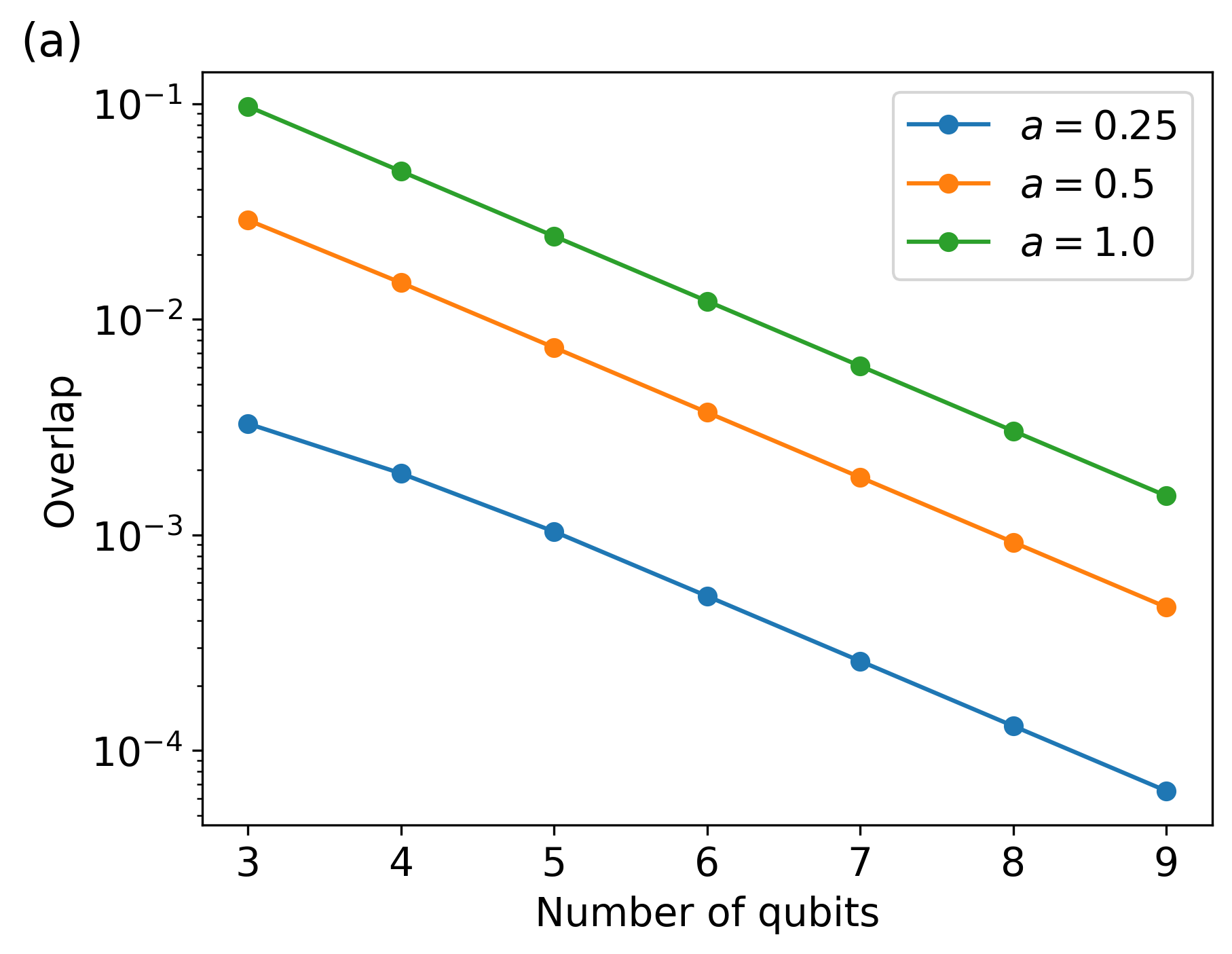}
  \end{minipage}
  \begin{minipage}[b]{0.85\hsize}
    \centering
    \includegraphics[width=0.85 \textwidth]{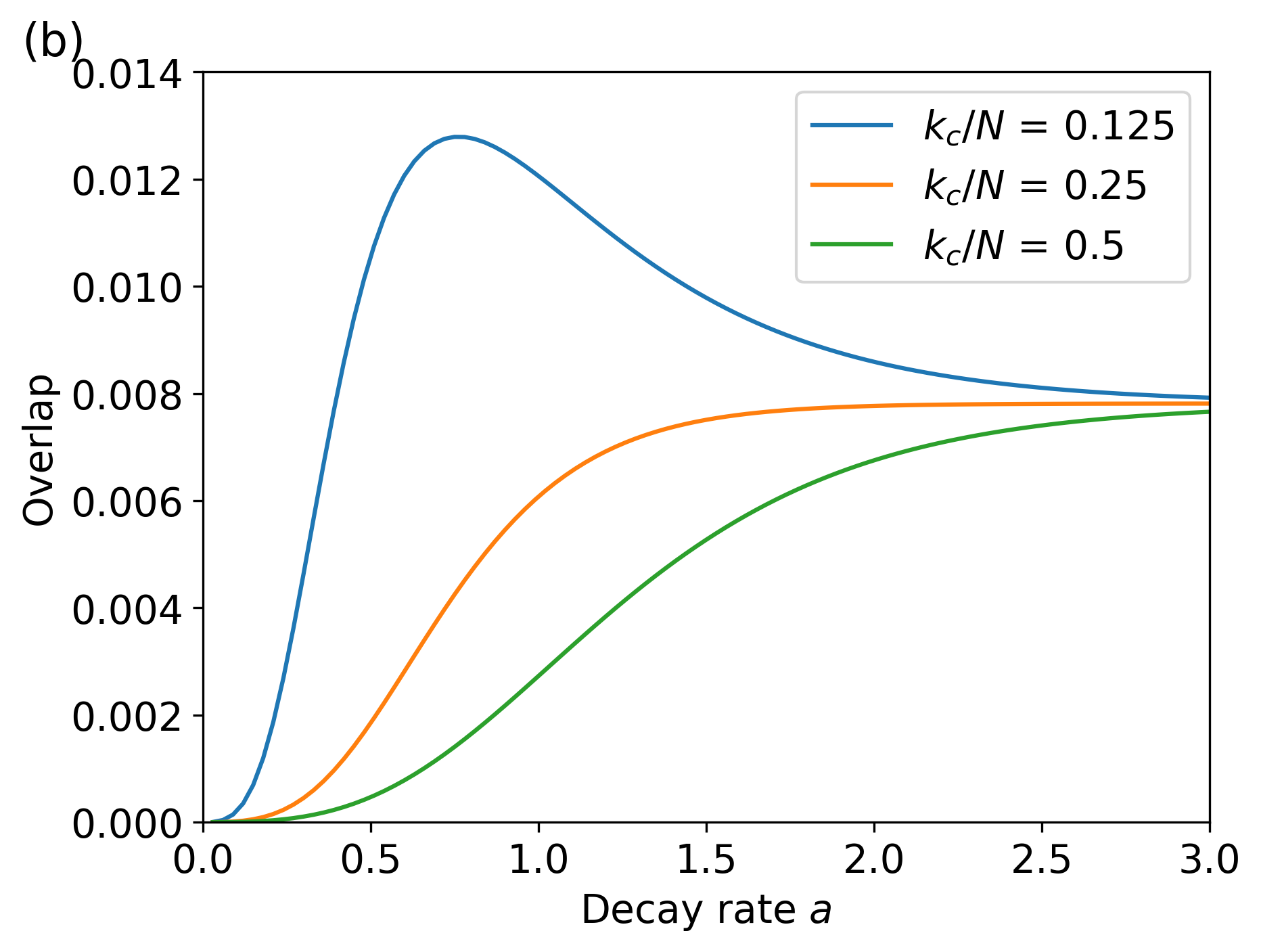}
  \end{minipage}
  \begin{minipage}[b]{0.85\hsize}
    \centering
    \includegraphics[width=0.85 \textwidth]{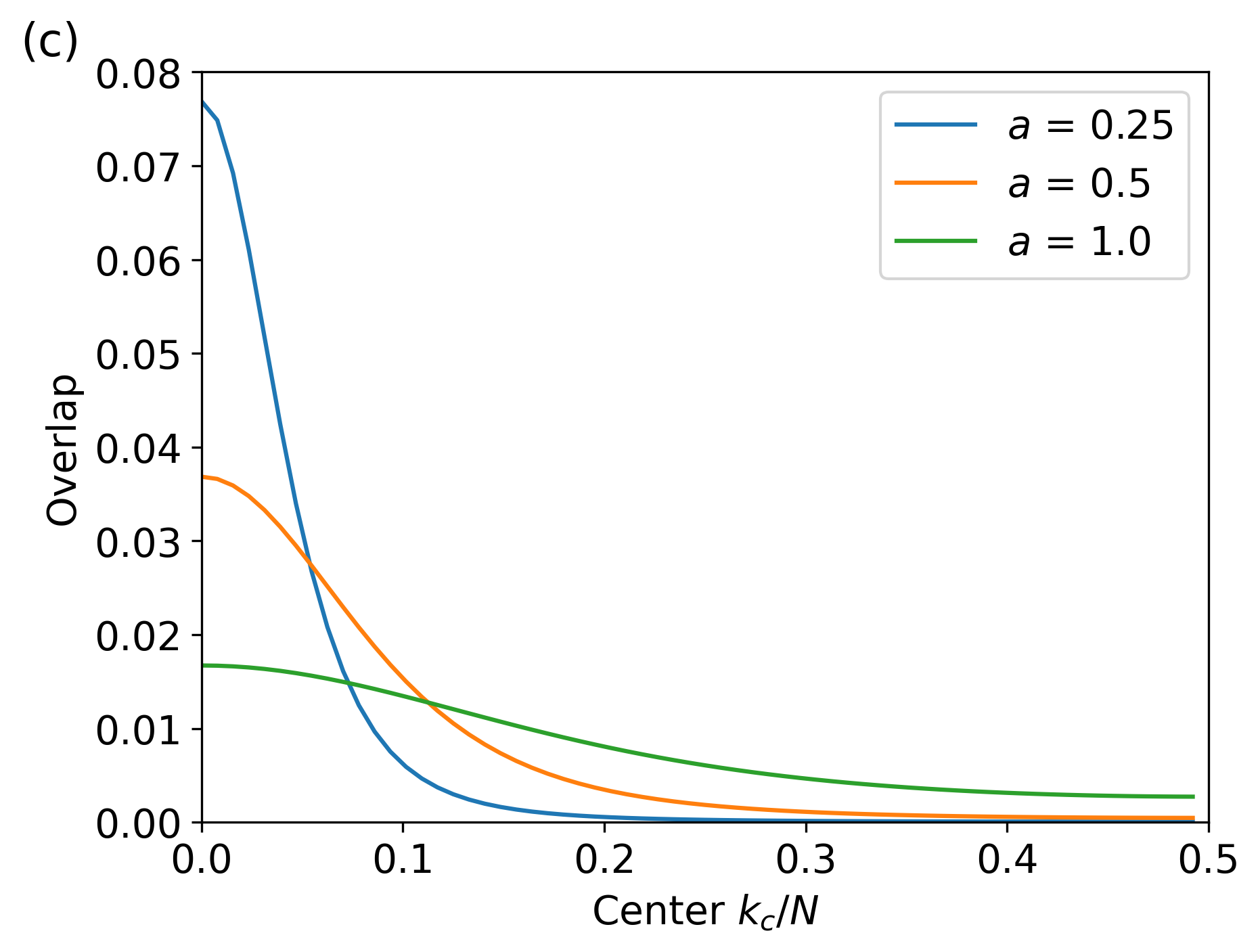}
  \end{minipage}
\caption{The overlap between two squared of LFs in Eq.~(\ref{eq:overlap2_two_lfs}) according to (a) the number of qubits, (b) the decay rate $a$,  and (c) peak center $k_c$. We used $k_c/N=0.25$ and $n=7$ in figure (a) and figures (b) and (c), respectively.}
\label{fig:overlap2_two_lfs}
\end{figure}

The dependency of the overlap between two squared of LFs on the number of qubits $n$, decay rate $a$, and peak center $k_c$ is plotted in Fig.~\ref{fig:overlap2_two_lfs}.
The overlap between two squared of LFs exponentially decreases as the number of qubits, as depicted in Fig.~\ref{fig:overlap2_two_lfs}(a). 
This phenomenon  can be more readily understood by considering the limit of a large decay rate $a$. In the limit, the LF state becomes a superposition state of equal weights. At this situation, the overlap is calculated as $\sum_{k=0}^{N-1} (1/N)\cdot (1/N) = 1/N$, thereby elucidating the reason for the exponential decay of the overlap with the number of qubits.
The overlap decreases monotonically except when $k_c/N=0.125$ as the decay rate decreases, as shown in Fig.~\ref{fig:overlap2_two_lfs}(b).
In the limit of a small decay rate, the LF state approaches a delta function, thereby resulting in an overlap that approaches zero. Therefore, the overlap exhibits a local maximum at $a=0.75$.
The overlap exhibits a monotonic decrease according to the peak center $k_c$, as shown in Fig~\ref{fig:overlap2_two_lfs}(c). The maximum value is larger when the decay rate is smaller.

\section{Details of QPE-LF}
\label{sect:comparison_qpe}
Spectral functions are commonly represented using Lorentzian broadening. To reproduce such broadening in QPE sampling, Ref.~\cite{Fomichev2024arXiv} proposed using a one-sided discrete exponential, or SF, state as the input state of the ancilla qubits. Unlike the periodic origin-centered SF state introduced in Eq.~(\ref{eq:slater_function_state}), the amplitudes of this state decay monotonically as $e^{-a\tau}$ over $\tau=0,\ldots,N-1$, without the symmetric branch associated with the periodic boundary condition. We denote its normalization constant by $\widetilde{C}_S(n,a)$ to distinguish it from the normalization constant $C_S(n,a)$ of the symmetric discrete SF state in Eq.~(\ref{eq:norm_slater_function_state}). The input state is given by
\begin{gather}
    |\psi^{\mathrm{SF}}_{\mathrm{in}} \rangle
    =
    \widetilde{C}_S(n, a) \sum_{\tau=0}^{N-1} e^{-a\tau}|\tau\rangle_{n}  ,
\end{gather}
where the normalization constant 
\begin{gather}
    \widetilde{C}_S(n, a)
    =
    \sqrt{
        \frac{1 - e^{-2a}}{1 - e^{-2aN}}
    } ,
\end{gather}
with positive decay rate $a>0$. In the $a\to 0$ limit,  $|\psi^{\mathrm{SF}}_{\mathrm{in}} \rangle$ becomes the input state of the original QPE. 
The state $|\psi^{\mathrm{SF}}_{\mathrm{in}} \rangle$ is generated by the product of $R_Y$ gates as \cite{Klco2020PRA}
\begin{gather}
    |\psi^{\mathrm{SF}}_{\mathrm{in}} \rangle
    =
    \bigotimes_{m=0}^{n - 1} 
    R_Y(2\theta_m) |0\rangle_{n}
\end{gather}
where
\begin{gather}
    \theta_m 
    \equiv
    \arctan \exp(-2^m a) .
\end{gather}
The function $\alpha$ is calculated as
\begin{gather}
\begin{aligned}
    \alpha^{\mathrm{LF}} (x)
    & =
    \frac{\widetilde{C}_S(n, a)}{\sqrt{N}}
    \sum_{\tau=0}^{N-1}
    e^{-a\tau}
    e^{2\pi i \tau x / N}
    \\ &=
    \frac{\widetilde{C}_S(n, a)}{\sqrt{N}}
    \frac{
        1 - e^{2\pi i x - a N}
    }{
        1 - e^{2\pi i x / N -a }
    }
    \\ & \approx
    \frac{\widetilde{C}_S(n, a)}{\sqrt{N}}
    \frac{1}{a - 2\pi i x / N}
\end{aligned}   
\end{gather}
and the absolute squared of $\alpha$ becomes
\begin{gather}
\begin{aligned}
    |\alpha^{\mathrm{LF}} (x)|^2
    &=
    \frac{\widetilde{C}_S^2(n, a)}{N}
    \left|
        \frac{
            1 - e^{2\pi i x - a N}
        }{
            1 - e^{2\pi i x / N -a }
        }
    \right|^2
    \\ & \approx
    \frac{\widetilde{C}_S^2(n, a)}{N}
    \left|
        \frac{1}{a - 2\pi i x / N}
    \right|^2
    \\ & =
    \frac{\widetilde{C}_S^2(n, a)}{N}
    \frac{1}{a^2 + 4\pi^2 x^2 /N^2}  .
\label{eq:alpha_slater}
\end{aligned}
\end{gather}
Here we used the following relation:
\begin{gather}
    \left|
        \sum_{\tau=0}^{N-1}
        e^{\tau (iy - \eta)}
    \right|^2
    \approx
    \left|
        \frac{1}{y + i\eta}
    \right|^2
    =
    \frac{1}{y^2 + \eta^2} .
\label{eq:appendix_alpha_slater_eq1}
\end{gather}
By introducing $\eta \equiv a N / 2\pi$, Eq.~(\ref{eq:appendix_alpha_slater_eq1}) follows
\begin{gather}
    |\alpha^{\mathrm{LF}}(x)|^2
    =
    \frac{\widetilde{C}_S^2(n, a) N}{(2\pi)^2}
    \frac{1}{x^2 + \eta^2}
\end{gather}
The limit of $x\to 0$ becomes
\begin{gather}
        |\alpha^{\mathrm{LF}} (x=0)|^2
        =
        \frac{\widetilde{C}_S^2(n, a)}{N a^2} .
\end{gather}
So, the maximum height is decreased as the decay rate $a$ increases.
Also, in limit of $a \to 0$, $|\alpha^{\mathrm{LF}}(x)|^2$ approaches $|\alpha(x)|^2$.
For large $N$ and fixed $\eta$, $\widetilde{C}_S^2$ scales as
\begin{gather}
\begin{aligned}
    \widetilde{C}_S^2\left(
    n, \frac{2\pi\eta}{N}
    \right)
    =
    \frac{
        1 - e^{-4\pi \eta / N }
    }{
        1 - e^{-4\pi \eta  }
    }
    \to
    O\left(\frac{\eta}{N}\right).
\end{aligned}
\end{gather}
The comparison of $|\alpha^{\mathrm{LF}} (x)|^2$ with that of the original QPE and the entanglement phase estimation is discussed in \cite{Nishi2026PRApplied}.

\section{Fidelity-based approach for spectra obtained by QPE-sampling}
\label{sect:appendix_fidelity_based_qpe}
This section presents a fidelity-based characterization of the histogram of QPE sampling.

By tracing out the system register of the QPE-output state in Eq.~(\ref{eq:qpe_output_re}), we obtain the reduced density matrix of the QPE register:
\begin{equation}
    \rho_{\mathrm{QPE}}
    \equiv
    \operatorname{Tr}_{s}
    \left[
        |\psi_{\mathrm{QPE}}\rangle
        \langle \psi_{\mathrm{QPE}}|
    \right]
    =
    \sum_{k=0}^{N-1}
    \left|\widetilde{c}_k\right|^2
    |k\rangle\langle k|,
    \label{eq:qpe_reduced_density}    
\end{equation}
where we have used
$\langle\widetilde{\phi}_k|\widetilde{\phi}_{k'}\rangle
=\delta_{k,k'}$, which holds in the ideal-QPE limit or when the QPE register is effectively dephased in the computational basis.

We define the fidelity between this reduced density matrix and the normalized LCNOF state in Eq.~\eqref{eq:mlf_state} as
\begin{align}
    F(\boldsymbol{d},\boldsymbol{a},\boldsymbol{k}_c)
    &\equiv
    \langle \psi_{\mathrm{LCNOF}}|
    \rho_{\mathrm{QPE}}
    | \psi_{\mathrm{LCNOF}} \rangle
    \nonumber\\
    &=
    \sum_{k=0}^{N-1}
    \left|\widetilde{c}_k\right|^2
    \left|
        \sum_{\ell=0}^{n_{\mathrm{bas}}-1}
        d_\ell
        L_{k-k_{c\ell}}(n,a_\ell)
    \right|^2.
    \label{eq:qpe_reduced_fidelity}
\end{align}
The fidelity can reach unity only when the reduced density matrix $\rho_{\mathrm{QPE}}$ is pure state and its corresponding pure state can be represented by the LCNOF state.

For fixed $\boldsymbol{a}$ and $\boldsymbol{k}_c$, maximizing this
fidelity under the normalization condition leads to the generalized
eigenvalue equation in Eq.~(\ref{ampl_encoding_of_GMO:eig_prob_for_lc_coeffs}), with
\begin{equation}
    G_{\ell,\ell'}
    =
    \langle L;a_\ell,k_{c\ell}|
    \rho_{\mathrm{QPE}}
    |L;a_{\ell'},k_{c\ell'}\rangle.
    \label{eq:qpe_reduced_overlap_matrix}
\end{equation}
The matrix elements $G_{\ell,\ell'}$ can be evaluated using the
circuit shown in Fig.~\ref{circuit:swap_test_qpe_fid}. The joint probability of observing the QPE register in $|0\rangle_n$ and the
ancilla qubit in $|j\rangle$ is
\begin{equation}
    P_j
    =
    \frac{
        G_{\ell,\ell}
        +
        G_{\ell',\ell'}
    }{4}
    +
    \frac{(-1)^j}{2}
    \operatorname{Re}
    \left[
        e^{i\phi}G_{\ell,\ell'}
    \right].
    \label{eq:qpe_reduced_overlap_probability}
\end{equation}
Thus, the real and imaginary parts of $G_{\ell,\ell'}$ can be obtained by changing $\phi$. For the real-valued LF states considered here, $G_{\ell,\ell'}$ is real, and only its real part needs to be evaluated.

\begin{figure}[h]
\centering
\mbox{ 
\Qcircuit @C=1em @R=1.6em{
\lstick{|0\rangle}                  & \qw                       & \qw      & \gate{H}  & \gate{Z(\phi)}                
& \ctrl{1}             & \ctrlo{1}            &  \gate{H} & \meter 
\\
\lstick{|0\rangle_{n}}              & \ustick{\otimes n} \qw  & {/} \qw  & \qw & \multigate{1}{\text{QPE}} 
& \gate{U_{\mathrm{LF}}^{(\ell) \dagger}} & \gate{U_{\mathrm{LF}}^{(\ell^{\prime}) \dagger}} & \qw       & \meter 
\\
\lstick{|\Psi_{\mathrm{in}}\rangle} & \ustick{\otimes n_s} \qw  & {/} \qw & \qw & \ghost{\text{QPE}}        
& \qw                  & \qw                  &  \qw      & \qw \\
}
} 
\caption{
Quantum circuit for evaluating the element of the matrix $G_{\ell, \ell^{\prime}}(\boldsymbol{a}, \boldsymbol{k}_c)$ in Eq.~(\ref{eq:qpe_reduced_overlap_matrix}).
The unitary operator $U_{\mathrm{LF}}^{(\ell)}$ generates the discrete LF state as $U_{\mathrm{LF}}^{(\ell)}|0\rangle_n = |L; a_{\ell}, k_{c\ell}\rangle$.
}
\label{circuit:swap_test_qpe_fid}
\end{figure}

\onecolumngrid 
\bibliography{ref}

\end{document}